\documentclass[lettersize,journal]{IEEEtran}
\usepackage{amsmath,amsfonts}
\usepackage{algorithmic}
\usepackage{array}
\usepackage[caption=false,font=normalsize,labelfont=sf,textfont=sf]{subfig}
\usepackage{textcomp}
\usepackage{stfloats}
\usepackage{hyperref}
\usepackage{url}
\usepackage{amsmath}
\usepackage{amssymb}
\usepackage{amsthm} 
\usepackage{xcolor}
\usepackage{booktabs, makecell, multirow, tabularx}

\usepackage[ruled, linesnumbered, noend]{algorithm2e}
\usepackage{verbatim}
\usepackage{graphicx}
\def\BibTeX{{\rm B\kern-.05em{\sc i\kern-.025em b}\kern-.08em
    T\kern-.1667em\lower.7ex\hbox{E}\kern-.125emX}}
\usepackage{balance}
\begin{document}
\title{EmInspector: Combating Backdoor Attacks in Federated Self-Supervised Learning Through Embedding Inspection}
\author{\IEEEauthorblockN{
Yuwen Qian, 
Shuchi Wu, 
Kang Wei, 
Ming Ding,
Di Xiao, 
Tao Xiang, 
Chuan Ma,
Song Guo, \emph{Fellow, IEEE}} 
}


\markboth{Journal of \LaTeX\ Class Files,~Vol.~18, No.~9, September~2020}%
{How to Use the IEEEtran \LaTeX \ Templates}

\maketitle

\begin{abstract}
Federated self-supervised learning (FSSL) has recently emerged as a promising paradigm that enables the exploitation of clients’ vast amounts of unlabeled data while preserving data privacy. While FSSL offers advantages, its susceptibility to backdoor attacks, a concern identified in traditional federated supervised learning (FSL), has not been investigated. To fill the research gap, we undertake a comprehensive investigation into a backdoor attack paradigm, where unscrupulous clients conspire to manipulate the global model, revealing the vulnerability of FSSL to such attacks. In FSL, backdoor attacks typically build a direct association between the backdoor trigger and the target label. In contrast, in FSSL, backdoor attacks aim to alter the global model's representation for images containing the attacker's specified trigger pattern in favor of the attacker's intended target class, which is less straightforward.  In this sense, we demonstrate that existing defenses are insufficient to mitigate the investigated backdoor attacks in FSSL, thus finding an effective defense mechanism is urgent. 
To tackle this issue, we dive into the fundamental mechanism of backdoor attacks on FSSL, proposing the \textit{Embedding Inspector (EmInspector)} that detects malicious clients by inspecting the embedding space of local models. 
In particular, EmInspector assesses the similarity of embeddings from different local models using a small set of inspection images (e.g., ten images of CIFAR100) without specific requirements on sample distribution or labels. We discover that embeddings from backdoored models tend to cluster together in the embedding space for a given inspection image. Evaluation results show that EmInspector can effectively mitigate backdoor attacks on FSSL across various adversary settings. Our code is avaliable at \href{https://github.com/ShuchiWu/EmInspector}{https://github.com/ShuchiWu/EmInspector}.
\end{abstract}

\begin{IEEEkeywords}
Federated learning, self-supervised learning, backdoor attack.
\end{IEEEkeywords}

\section{Introduction}
\IEEEPARstart{A}{lthough} self-supervised learning (SSL) frees model training from the burdens of costly data annotation efforts \cite{liu2021self}, it demands a massive amount of data to achieve comparable performance to its supervised counterpart, e.g., 400 million image-text pairs for training the CLIP \cite{radford2021learning} and one billion noisy image-text pairs for training the ALIGN \cite{jia2021scaling}. Such a massive volume of data can be overwhelming for individuals or organizations and could become an insurmountable obstacle for model training in certain fields where data is scarce. In this regard, federated SSL (FSSL) employs a distributed learning approach and offers an alternative to address the data provision issue. This is because contemporary edge devices have access to a wealth of data suitable for learning models \cite{mcmahan2017communication}. Various research works of FSSL on practical usage such as audio recognition \cite{tsouvalas2022federated}, video understanding \cite{rehman2022federated} and disease diagnosis \cite{wu2022federated}, have emerged recently, showcasing its immense potential.

However, as the traditional federated supervised learning (FSL) paradigm is vulnerable to backdoor attacks due to its decentralized nature \cite{sun2019can,bagdasaryan2020how}, this vulnerability may also be inherited in FSSL. According to our survey, prior research has mainly focused on the vulnerability of traditional FSL \cite{gong2022coordinated, Xie2020DBA, sun2019can, bagdasaryan2020how, wang2020attack} or centralized SSL frameworks \cite{saha2022backdoor, jia2022badencoder} while lacking in treating in FSSL. 

\begin{figure}[t]
    \centering
    \includegraphics[width=0.85\linewidth]{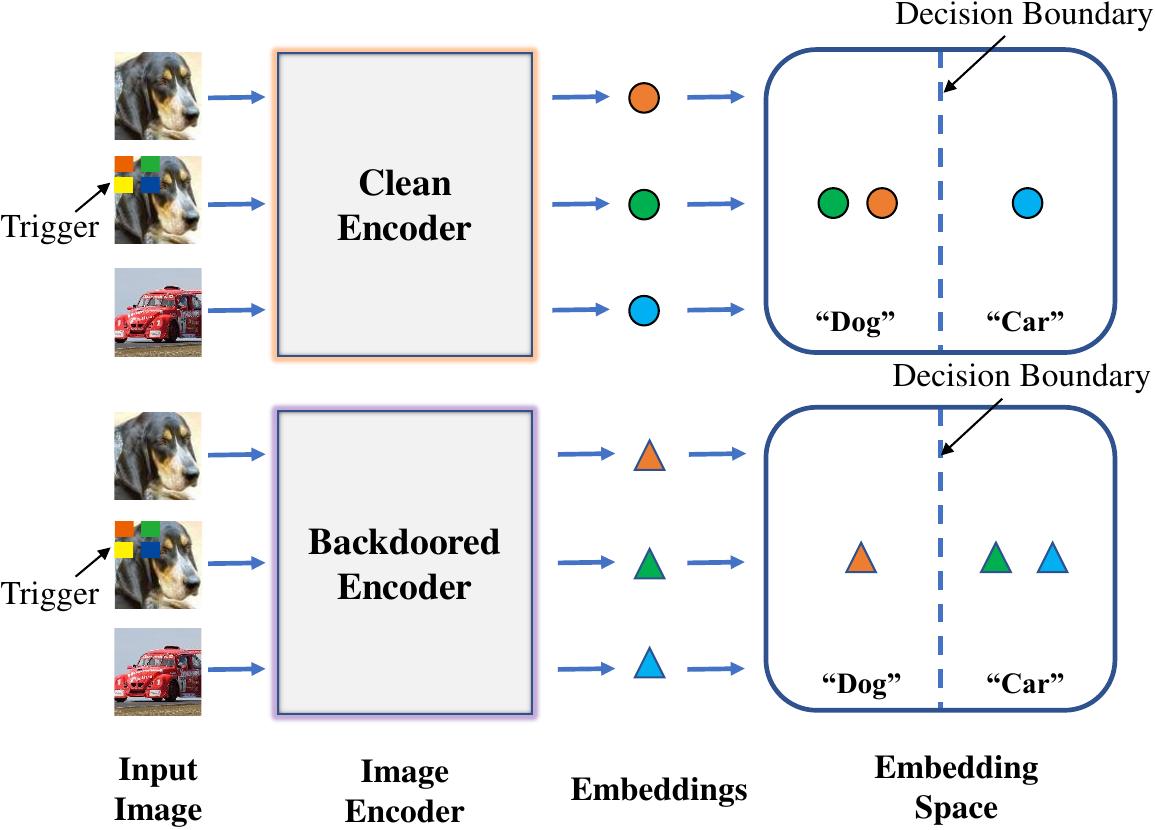}
    \caption{An effective backdoor attack in SSL should make the encoder pull the embedding of the image with a trigger to the target class (below, take ``car'' as the target class for instance), while the clean encoder performs normally (above).}
    \label{fig:backdoor on ssl}
\end{figure}

\noindent
\textbf{Distinctions of backdoor attacks in FSSL and FSL.}
In FSL, malicious clients execute backdoor attacks by uploading crafted models trained on mismatched data-label (i.e., the target label) pairs with a trigger. The aim is to mislead the global model to predict the attacker's desired label for inputs embedded with the attacker's chosen trigger. In FSSL, each trained model only outputs embeddings (features) of input data, and the predicting process relies on downstream classifiers. Thus, for a successful backdoor attack, malicious clients should modify their local models to generate similar embeddings for inputs containing the attacker's chosen trigger and twist these embeddings to the attacker's target class \cite{saha2022backdoor, jia2022badencoder}, as depicted in Fig. \ref{fig:backdoor on ssl}. As a result, any downstream classifier built based on the frozen backdoored global model as a backbone will misclassify inputs with a similar trigger into the same target class. In conclusion, backdoor attacks in FSL build a direct association between the backdoor trigger and the target label in the low-dimensional label space (e.g., 10 in CIFAR10) while in FSSL, backdoors are embedded in the high-dimensional embedding space (e.g., 512 of ResNet18), which is less explicit and more challenging to detect.

\noindent
\textbf{Infeasibilities of existing defenses.}
In the literature, there have been numerous defenses for safeguarding FL against model poisoning attacks like untargeted attacks that aim to degrade the model performance \cite{Dong2018Byzantine, blanchard2017machine, Cao2021FLTrust, pillutla2022robust} and backdoor attacks \cite{nguyen2022flame, fung2020limitations, wang2022rflbat, wang2022flare, andreina2021baffle}. Most of them are based on statistical characteristics of uploaded local model parameters/gradients \cite{blanchard2017machine, Dong2018Byzantine, Cao2021FLTrust, pillutla2022robust, nguyen2022flame, fung2020limitations, wang2022rflbat}, aiming to filter outliers. Besides inspecting model parameters directly, some others use a set of inspection data to acquire feedback (e.g., labels and embeddings) from uploaded local models to detect whether they are backdoored \cite{andreina2021baffle, wang2022flare}. 
\textbf{Both the existing statistical-based and feedback-based defense strategies share a common underlying principle: they aim to identify and discard potentially malicious models by assessing their dissimilarity to the majority of models recognized as benign.} However, thanks to the high-dimensional outputs in FSSL, backdoored models can be optimized with a loss constraint (Eq. \ref{eq:3}) to have very similar outputs on clean samples with benign models. 
This practice also suppresses the deviation in model parameters between the backdoored model and its benign counterpart to a teeny-tiny level.
\textbf{Furthermore, defeating the intuition proposed in the literature of FSL \cite{wang2022flare}, we discover that one backdoored local model in FSSL is prone to generate more similar embeddings (i.e., non-outliers) with other models when inputting identical images.} Therefore, existing defenses designed based on outlier detection (both statistical-based and feedback-based) could fail, as we demonstrated in Sec. \ref{sec: experimental results}.

\noindent
\textbf{Our works and contributions.} To the best of our knowledge, we are the first to conduct a comprehensive exploration of backdoor attacks on FSSL to fill this research gap. 
Generally, backdoor attacks on FSSL can be divided into two paradigms depending on the attacker's capability: one in which the attacker can only poison a limited number of images in clients' training datasets, and the other in which the attacker has full control over a certain number of malicious clients to manipulate the loss function, hyper-parameters, training data, optimizer, and training process arbitrarily. According to the experimental results (refer Fig. \ref{fig:cvpr_attack} in Appendix \ref{appendix:experiment results}), we observe that the first data-poisoning-based attack paradigm has minimal effects on FSSL owing to the strict requirements for its success and the damping effect caused by data augmentations used by SSL and global aggregation. Therefore, we primarily focus on the second paradigm, which poses a more substantial threat to FSSL systems. We relegate the detailed description of the first paradigm to Appendix \ref{appendix: Data-poisoning-based backdoor attack}.
Moreover, we demonstrate that existing defenses are inadequate in mitigating the investigated backdoor attacks in FSSL. To counter the threat of backdoor attacks in FSSL, we dive into the fundamental working mechanism of backdoor attacks on FSSL and then elaborate on the idea of embedding inspection to propose a novel defense mechanism for FSSL against backdoor attacks, i.e. Embedding Inspector (EmInspector). Our EmInspector mechanism is predicated on a crucial observation, i.e., for a given image, embeddings generated by backdoored local models tend to cluster together and stay within the region of embeddings produced by benign local models in the embedding space (refer Fig. \ref{fig: embedding space}). Extensive experiments demonstrate that EmInspector only requires minimal inspection images (e.g., ten samples are enough in experiments with CIFAR100) to accurately identify backdoored models, without specific requirements on sample distribution or labels. Our contributions can be summarized as follows:

\begin{itemize}
\item[$\bullet$] We show that model-poisoning-based backdoor attacks are highly effective against FSSL and can be launched at any round while achieving a high attack success rate without degrading the primary task accuracy, which contravenes previous understanding, thereby highlighting the severity of the problem. 
\item[$\bullet$] To defend against backdoor attacks in FSSL, we propose EmInspector via inspecting abnormal embeddings by feeding inspection images to uploaded local models. EmInspector is the first defense mechanism to tackle this security problem that requires only minimal inspection images to accurately detect malicious clients.
\item[$\bullet$] 
Our extensive evaluations across various settings (e.g., datasets, training data distribution (non-i.i.d or i.i.d), adversary settings, networks) show that EmInspector can dynamically detect malicious clients without requiring prior knowledge about the adversary settings while exhibiting high effectiveness. 
\end{itemize}

\section{Preliminaries} \label{sec: related work}
\subsection{Federated Self-Supervised Learning}
An FSSL system typically consists of two components: a pretext task at the client side, designed to learn representations from unlabelled data, and an aggregation algorithm at the server side to aggregate local models uploaded by clients. 
For pretext tasks, a commonly used approach is contrastive learning, as demonstrated by methods like SimCLR \cite{chen2020simple} and MoCo \cite{he2020momentum}. In contrastive learning, two different augmentations of the same image are considered as a positive pair, and the model aims to make their representations similar. On the other hand, two different images are treated as a negative pair, and the model strives to make their representations dissimilar. This approach enables the model to learn meaningful representations that capture the underlying structure of the data.
For aggregation, FedAVG \cite{mcmahan2017communication} is the most widely-used method which uses the average of all uploaded models as the new global model. Moreover, as data held by different clients usually have various distributions, model consistency of the aggregated model has gained much popularity and motivated various research \cite{li2022federated, Disha2022Federated, zhuang2022divergenceaware, Liu2022Feature}. However, existing works focus on addressing the data heterogeneity caused by the distributed setting for FL regardless of possible backdoor risks.

\subsection{Backdoor Attacks in FL \& SSL}
The objective of a backdoor attack is to modify the model so that it predicts the attacker's desired labels for inputs embedded with the attacker's chosen trigger pattern, while still being able to predict correct labels for clean inputs with no significant decrease in accuracy.
Backdoor attacks on FL have been recently studied, to manipulate a subset of training data of some local clients such that ML models trained on poisoned datasets with the crafted trigger will be vulnerable to test data with a similar trigger embedded, as shown in \cite{bagdasaryan2020how, Xie2020DBA, gong2022coordinated}.
Bagdasaryan \textit{et al.} \cite{bagdasaryan2020how} utilized a global trigger pattern (image patch chosen and shared by all malicious clients) to poison the training data, while Xie \textit{et al.} \cite{Xie2020DBA} proposed an attacking paradigm where malicious clients work in collusion via multiple distributed local patterns, which has shown a higher attack success rate, faster convergence and better resiliency than using a single global pattern.

In terms of SSL, backdoor attacks aim to inject backdoors into a pre-trained image encoder so that when a downstream classifier is built based on the backdoored encoder, it will predict the attacker's desired target labels for inputs with the attacker's chosen trigger \cite{saha2022backdoor,jia2022badencoder,li2022demystifying}.
For the attack, Saha \textit{et al.} \cite{saha2022backdoor} proposed to poison the training dataset of the victim.
BadEncoder \cite{jia2022badencoder} assumes access to a clean pre-trained image encoder and a set of unlabeled images.
By fine-tuning the pre-trained encoder with an elaborately designed loss function and a reference image of the attacker's target class, BadEncoder aims to mislead the encoder to recognize data containing the specific trigger pattern as the target class.

However, prior research has solely emphasized the susceptibilities of conventional FL or centralized SSL frameworks, without exploring the vulnerabilities present in FSSL. 

\subsection{Defense Mechanisms for FL}
In this paper, we consider advanced defenses against both untargeted attacks that aim to degrade the model performance \cite{shejwalkar2021manipulating} and backdoor attacks as our baselines.

\noindent
\textbf{Defenses against untargeted attacks.} Krum \cite{blanchard2017machine} selects a single model that has the smallest Euclidean distance with others as the new global model. While Krum focuses on choosing a single model, Trimmed-mean \cite{Dong2018Byzantine} is a coordinate-wise aggregation algorithm that aggregates each model parameter independently based on the trimmed mean of each parameter. Despite only relying on the uploaded models, FLTrust \cite{Cao2021FLTrust} bootstraps trust by maintaining a model on the server side.

\noindent
\textbf{Defenses against backdoor attacks.}
In the literature, most current defenses against backdoor attacks in FL are designed based on inspecting local models' parameters or gradients directly to detect outlier models: Deepsight \cite{rieger2022deepsight} performs deep model inspection for each model by checking its Normalized Update Energies (NEUPs) and Division Differences (DDifs). Foolsgold \cite{Fung2020} punishes clients that have historical updates with high pairwise cosine similarities by a low learning rate. FLAME \cite{nguyen2022flame} bounds the added noise of differential privacy defenses to mitigate the negative impact on primary tasks. RFLBAT \cite{wang2022rflbat} applies Principal Component Analysis (PCA) to separate malicious models from benign models in a low dimensional space. 
Most related to our work that detects malicious clients through inspecting the embedding space is Flare \cite{wang2022flare}, which is designed based on an observation that embeddings from backdoored models are outliers among those from benign models. 

Despite providing enhanced security for FL, our experiments demonstrate that existing methods are inadequate in defeating backdoor attacks in FSSL.

\begin{figure*}[t]
\centering
\includegraphics[width=0.8\textwidth]{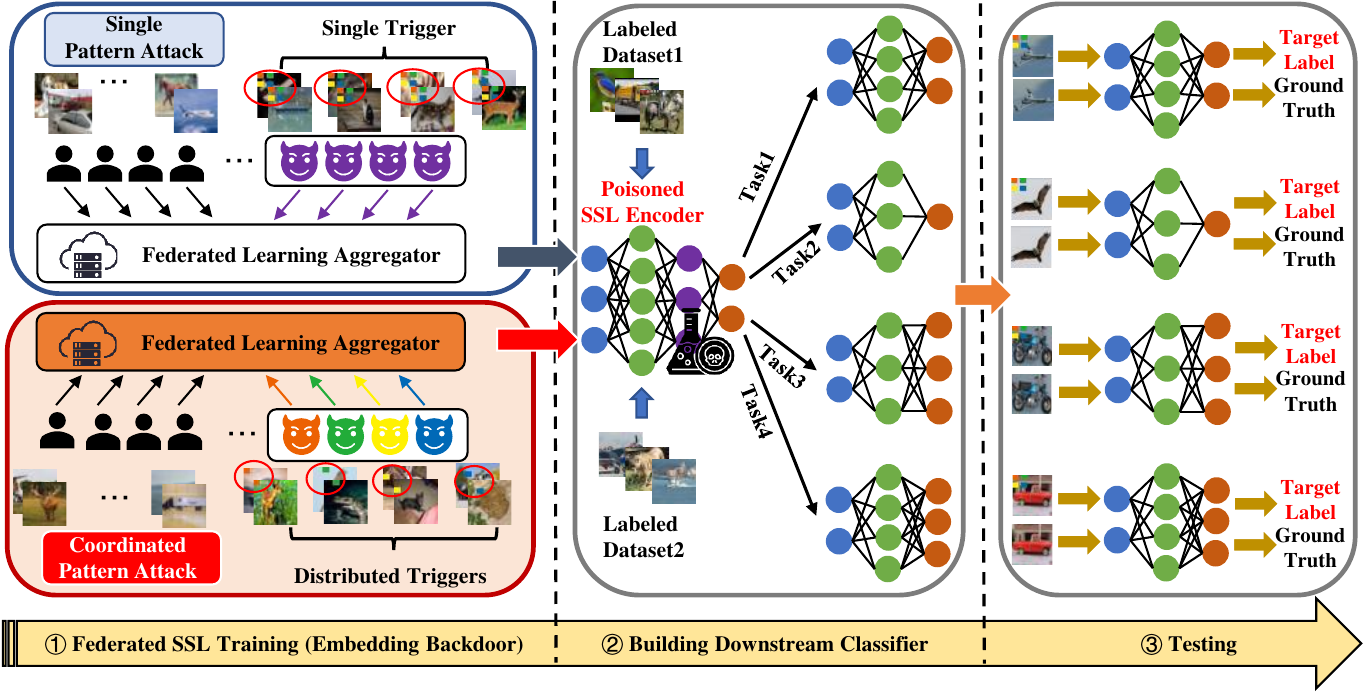}
\caption{Overview of backdoor attacks on FSSL with two different collusion methods. In the single pattern attack, all malicious clients use the same trigger pattern for attacking, while in the coordinated pattern attack, the attacker distributes a unique trigger for each malicious client and uses an assembled pattern of them to trigger the backdoor embedded in the aggregated model.}
\label{fig:attack_diagram}
\end{figure*}

\section{System Formulation} \label{sec: system setup}
\subsection{System Model of FSSL}
We consider a typical FSSL setting in which $N$ local clients collaboratively train a global SSL model.
In each communication round, each client uses its unlabelled local dataset $\mathcal{D}_{i} = \{ \boldsymbol{x}_{1},\ldots,\boldsymbol{x}_{\vert \mathcal{D}_{i}\vert}\}$, $i \in \mathcal{N} = \{1, 2,\cdots, N\}$ to train a local SSL model. 
Specifically, each communication round of FSSL consists of the following three steps:
\\\textbf{Step 1: Model broadcasting.} At the beginning of the $t$-th communication round, the server broadcasts the current global SSL model $\boldsymbol{\theta}^{t}$ to the selected clients.
\\\textbf{Step 2: Local training.} A client $i$ trains a local SSL model via fine-tuning the received global SSL model with its unlabeled local dataset $\mathcal{D}_i$. In particular, the client initializes its local model to the received global SSL model $\boldsymbol{\theta}^{t}$, and then updates the local model with an optimizer for $\tau$ local epochs. 
After training, each trained local SSL model $\boldsymbol{\theta}_i^{t,\tau}$ will be sent to the central cloud server.
\\\textbf{Step 3: Global aggregation.} The server obtains a new global model $\boldsymbol{\theta}^{t+1}$ via aggregating received local models. FedAVG \cite{mcmahan2017communication} is used for aggregation by default , i.e., $\boldsymbol{\theta}^{t+1} =  \sum_{i\in \mathcal{S}^{t}}p_{i}^{t}\boldsymbol{\theta}_i^{t,\tau}$, where $\mathcal{S}^{t}$ is the set of selected clients, and $p_{i}^{t} = \vert \mathcal D_i\vert/\sum_{i\in \mathcal{S}^{t}}{\vert \mathcal D_{i}\vert}\geq 0$. 

After finishing the training process, we froze the parameters of the global model, using it to build a downstream predictor with a few labeled training data.

\subsection{Attacker's Objectives and Capability}
We extend previous works\cite{jia2022badencoder, Xie2020DBA} in the federated setting to evaluate the vulnerability of the realistic FSSL system against backdoor attacks in our paper. 
We summarize the attacker's objectives and capability as follows.

\noindent
\textbf{Attacker's objective.} We assume that there exist several malicious clients aiming to inject backdoors into the global SSL model so that when a downstream classifier is built based on the poisoned global model, it will misclassify any image with the attacker's chosen trigger pattern to the attacker's target class. Also, malicious clients expect to maintain the accuracy of the primary task for stealthy purposes. 
Specifically, malicious clients may backdoor their models and upload these corrupt results to the server.

\noindent
\textbf{Attacker's capability.}
We consider that malicious clients have knowledge of the global model's architecture and can manipulate their local datasets, redesign the loss function, and alter the model training process arbitrarily. Moreover, malicious clients can collude with each other to enhance the effect of the backdoor attack, as we will interpret in detail later.  
However, malicious clients are largely unaware of other details of the FSSL system, including models from benign clients and the aggregation rule of the server.

\section{Backdoor Attacks on FSSL} \label{sec: framework of backdoor}
In each FSSL epoch, malicious and benign clients share backdoored and benign models respectively, and then the server updates the global model according to the aggregation rule. 
We consider an attacker to have complete control over $M$ malicious clients that work in collusion with each other and upload backdoored local models to the server.
The proportion of malicious clients is denoted as $p = \frac{M}{N}$, and we make the assumption that malicious clients always comprise the minority among all local clients, i.e., $p < 50\%$, otherwise, the system is worthless.
The set of malicious clients can be expressed as $\mathcal{M}$ and it satisfies $\vert \mathcal{M}\vert =M$ and $\mathcal{M} \subset \mathcal{N}$.
Next, we will introduce backdoor attacks on FSSL from three aspects: 1) injecting triggers, 2) collusion method, and 3) attack timing.

\noindent
\textbf{Injecting triggers.}
Specifically, each malicious client needs to inject triggers into its local model and keep them stealthy.
To achieve this objective, the first step for malicious clients is to generate similar feature vectors for data containing triggers and data about the target class (designated as reference data).
Moreover, to increase the potency and efficiency of the attack, they also ensure that the backdoored model accurately identifies the reference data by guiding it to produce a similar feature vector for the reference data with the clean model, as explained in \cite{jia2022badencoder}. 
Therefore, for $i' \in \mathcal{M}$, we can formulate this goal as $\mathcal{L}_{i',1}$, i.e.,
\begin{equation} \label{eq:2}
\begin{aligned}
\mathcal{L}_{i',1} &= - \frac{1}{{\left| {\mathcal{D}_{i'}} \right|}}{\sum_{\boldsymbol{x} \in \mathcal{D}_{i'}}s\left(f\left(\boldsymbol{x} \oplus \boldsymbol{e}_{i'}, \boldsymbol{\tilde{\theta}}_{i'}\right), f\left(\boldsymbol{x}_{\text{target}}, \boldsymbol{\tilde{\theta}}_{i'}\right)\right)} \\
 &\qquad \qquad  - s\left(f\left(\boldsymbol{x_{\text{target}}},  \boldsymbol{\tilde{\theta}}_{i'}\right), f\left(\boldsymbol{x}_{\text{target}}, \boldsymbol{\theta}\right)\right),
\end{aligned}
\end{equation}
where $s(\cdot,\cdot)$ measures the similarity between two feature vectors, $f$ is the encoder, $f(\cdot,\boldsymbol{\tilde{\theta}}_{i'})$ is the output feature vector from the SSL model with parameter $\boldsymbol{\tilde{\theta}}_{i'}$, $\boldsymbol{\tilde{\theta}}_{i'}$ is the backdoored parameter to be optimized, $\boldsymbol{\theta}$ is a clean model parameter, $\boldsymbol{x} \oplus \boldsymbol{e}_{i'}$ represents a clean image $\boldsymbol{x}$ embedded with a trigger pattern $\boldsymbol{e}_{i'}$ and $\boldsymbol{x}_{\text{target}}$ represents a reference image of the target class. Formally, we use cosine similarity for $s(\cdot,\cdot)$ and use the clean global model from the previous round of launching attacks for $\boldsymbol{\theta}$. For simplicity, we will use the initial clean global model in the subsequent description of this paper for $\boldsymbol{\theta}$.

To keep the attack stealthy, malicious clients should not degrade the main task performance, for which they need to ensure the backdoored model generates similar feature vectors for clean inputs to a clean model.
This goal can be formulated as $\mathcal{L}_{i',2}$, i.e.,
\begin{equation} \label{eq:3}
    \mathcal{L}_{i',2} = - \frac{1}{{\left| {\mathcal{D}_{i'}} \right|}}{\sum_{\boldsymbol{x} \in \mathcal{D}_{i'}}s\left(f\left(\boldsymbol{x}, \boldsymbol{\tilde{\theta}}_{i'}\right), f\left(\boldsymbol{x}, \boldsymbol{\theta}\right)\right)},
\end{equation}
Combining $\mathcal{L}_{i',1}$ and $\mathcal{L}_{i',2}$, the goal of injecting triggers can be expressed as the following optimization problem:
\begin{equation}\label{eq:trade off}
    \underset{\boldsymbol{\tilde{\theta}}_{i'}}{{\arg\min} \, \mathcal{L}_{i'}} = \lambda_1 \cdot \mathcal{L}_{i',1}\left(\boldsymbol{\tilde{\theta}}_{i'}\right) + \lambda_2 \cdot \mathcal{L}_{i',2}\left(\boldsymbol{\tilde{\theta}}_{i'}\right),
\end{equation}
where $\lambda_1$ and $\lambda_2$ are preset weights to adjust the attack degree.
In practice, there is a tradeoff between the two values, i.e., a more aggressive attack or a stealthier one \cite{jia2022badencoder}.

\noindent
\textbf{Collusion method.}
We develop the collusion method based on the trigger selection.
A naive collusion method is that all malicious clients use the same trigger pattern $\boldsymbol{e}$ to train backdoored models.
Via this method, the trigger pattern $\boldsymbol{e}$ will be injected into all models from malicious clients, which can enhance the effect of this trigger pattern on the global model.
We refer to this method as \textit{single pattern attack}.
In contrast, each malicious client may use a unique local trigger pattern $\boldsymbol{e}_i, i = 1, 2, \cdots, M$, similar to \cite{Xie2020DBA}. The backdoor embedded in the global model can be triggered by the assembled pattern of the local triggers, i.e., $\bigcup_{i=1}^M \boldsymbol{e}_i$. This assembled pattern is denoted by the global trigger $\boldsymbol{e}_{\text{glob}}$. 
We refer to this method as \textit{coordinated pattern attack}.
For fair comparison, we set $\boldsymbol{e}=\boldsymbol{e}_{\text{glob}}$ in experiments. 
Fig. \ref{fig:attack_diagram} illustrates an overview of the two collusion methods for backdoor attacks on an FSSL system. Briefly speaking, malicious clients first work in collusion to backdoor the global SSL model, and the learned representation will transfer to downstream supervised tasks. We will make a comparison between the two collusion methods and demonstrate the effectiveness of EmInspector against both of them in Sec. \ref{eval: defense}.

\noindent
\textbf{Attack timing.}
In practice, malicious clients can launch attacks in any round of FSSL. 
However, they are more likely to launch attacks when the global model is close to convergence to reduce the negative effect on the performance of primary tasks, as pointed out by Xie \textit{et al.} \cite{Xie2020DBA}.
In addition, Bagdasaryan \textit{et al.} \cite{bagdasaryan2020how} suggests that initializing the attack when the global model is close to convergence would make the attack more efficient. Therefore, we begin attacking when the global model is close to convergence by default. We also conduct an ablation experiment to verify the impact of the attack timing and the result is presented in Sec. \ref{eval: attack}.

The algorithm of backdoor attacks on FSSL is summarized in algorithm \ref{alg:active attack} in Appendix \ref{appendix: alg1}.
We find that this attack paradigm can achieve backdoor objectives by fine-tuning model parameters to an extremely small extent in FSSL systems.
This makes backdoored and benign models of high similarity, causing the failures of various existing defense mechanisms, which we will discuss in Section \ref{eval: defense}. Therefore, it is necessary to develop a defense mechanism that can mitigate backdoor attacks effectively while maintaining the performance of the model on primary tasks in FSSL systems.

\begin{figure}[t]
    \centering
    \includegraphics[width = \linewidth]{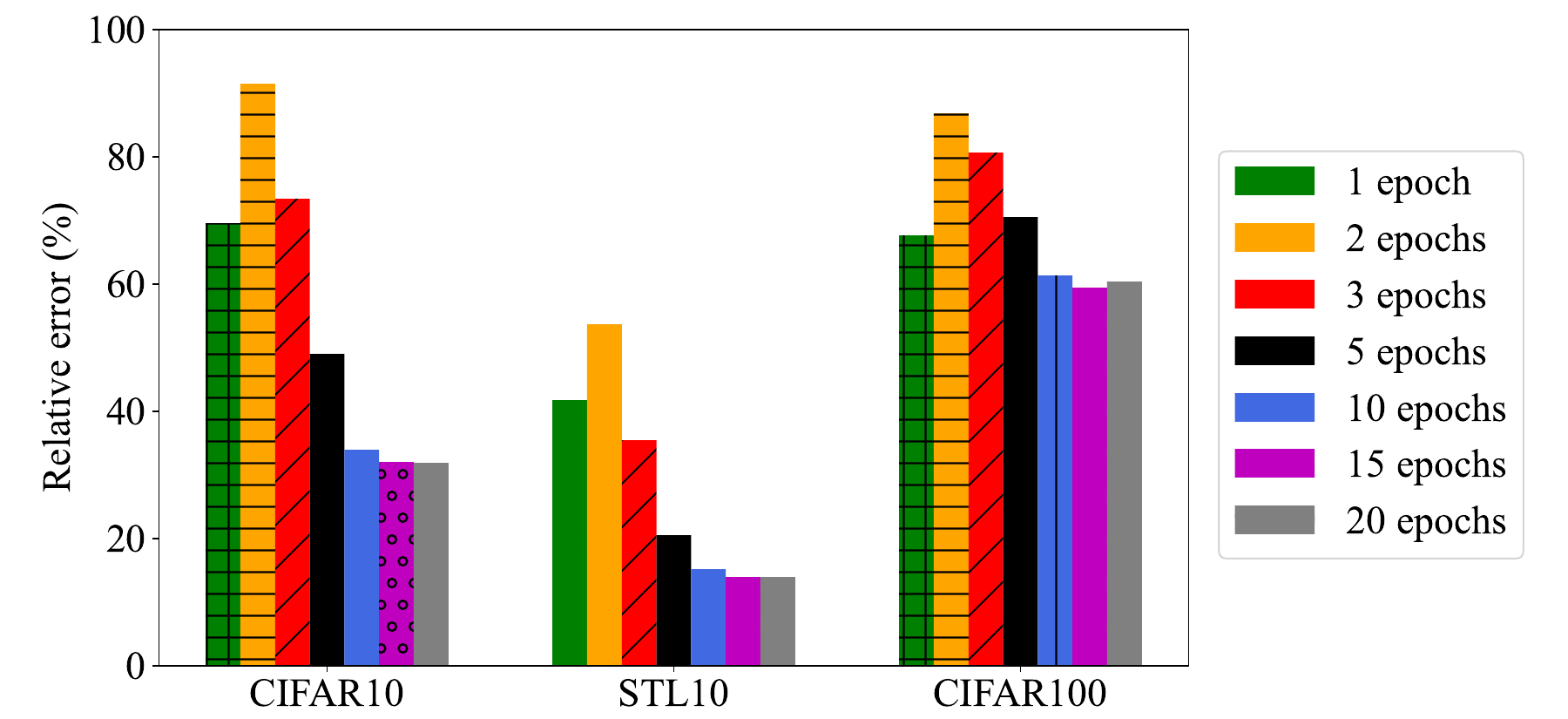}
        \caption{Relative errors of the gap between the target class and other classes measured by a backdoored and a benign model. Epochs for injecting backdoors vary from 1 to 20.}
    \label{fig:observation}
\end{figure}

\section{Design of EmInspector} \label{sec: design of embinspector}
\subsection{Key observations}
In FSSL systems with adversary settings, benign clients use unpoisoned local datasets to train their local models standardly, while malicious clients aim to achieve two goals, i.e., injecting backdoors into the global SSL model and keeping its performance on primary tasks as well as a benign one.
Under ideal conditions, the malicious can inject backdoors imperceptibly. However, recall for the loss function $\mathcal{L}_{i',1}$, we assume that the backdoored feature extractor will also explore information from the clean part of poisoned images (except triggers) as it cannot pay all attentions to the trigger, which makes $f(\boldsymbol{x}, \boldsymbol{\tilde{\theta}}_{i'})$ and $f(\boldsymbol{x}_{\text{target}}, \boldsymbol{\tilde{\theta}}_{i'})$ more similar than before.

To further validate this assumption, an experiment is carried out on CIFAR10 which involves using a backdoored model and a clean model to assess the similarity (gap) between the target class and other classes and compute the relative error. 
The backdoored models are trained for various epochs with an attack degree of $1:1$ (i.e., $\lambda_1:\lambda_2 = 1:1$) to observe the variation of the gap. The experimental details are provided in Appendix \ref{appendix:measuring the gap}, and the results are presented in Fig. \ref{fig:observation}, which shows that the backdoor does minimize the gap between the target class and other classes in the embedding space. Moreover, we can observe that the gap relative error increases first and then decreases to a stable value. This can be explained by the fact that the model initially cannot give much weightage to the trigger, and thus tends to rely more on the clean part of the poisoned images in the beginning, to explore the information. Along with the training process, the model will pay more attention to the trigger, and thus the gap relative error will decrease. 

\begin{figure}[t]
    \centering
    \includegraphics[width = \linewidth]{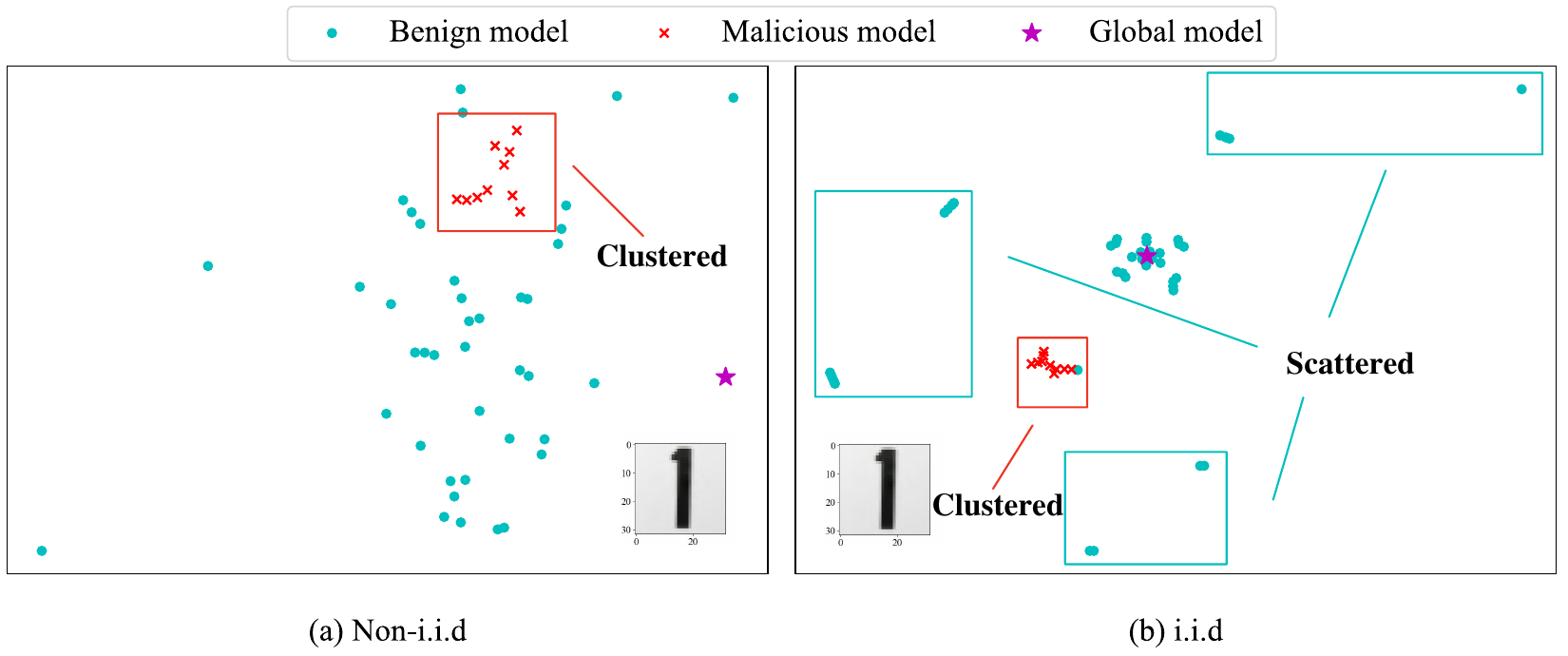}
    \caption{t-SNE plots of the embedding space. Embeddings are generated by 10 backdoored models, 40 benign models, and the current global model. (a) and (b) are plotted under a non-i.i.d and an i.i.d data setting, respectively. The input image is shown in the bottom right/left corner in (a)/(b).}
    \label{fig: embedding space}
\end{figure}

To further analyze the difference of embeddings from benign and backdoored models in a global perspective, in Fig. \ref{fig: embedding space}, we use t-SNE \cite{van2008visualizing} to visualize the embeddings of an image ``digit one'' (illustrated in the bottom right/left corner of (a)/(b) of Fig. \ref{fig: embedding space}) generated by 40 benign models, 10 backdoored models (trained for 5 epochs with 1,000 images), and the current global model that trained under a non-i.i.d data setting (Fig. \ref{fig: embedding space} (a)) or an i.i.d data setting (Fig. \ref{fig: embedding space} (b)).
From Fig. \ref{fig: embedding space}, three observations can be drawn: 
\\
$\bullet$ \textbf{Backdoors attract all embeddings to the target class.} 
In Fig. \ref{fig: embedding space}, we can observe that backdoored models tend to have densely clustered embeddings in the embedding space. This is attributed to that backdoored models push the embeddings of all images closer to the reference ones, as we discussed above. In detail, consider $\theta_i$ and $\theta_j$ are two clean SSL models and we assume $s(f(\boldsymbol{x}, \theta_i), f(\boldsymbol{x}, \theta_j)) \geq \sigma$. After $\theta_i$ and $\theta_j$ being backdoored, for the new lower bound $\sigma^{'}$, we have $\sigma^{'} \geq \sigma$. A brief proof is provided for this observation:
\vspace{-3mm}
\begin{proof}
Due to $\arccos (s(\boldsymbol{a}, \boldsymbol{c}))\leq\arccos (s(\boldsymbol{a}, \boldsymbol{b})) + \arccos (s(\boldsymbol{b}, \boldsymbol{c}))$, we have 
\vspace{-2.5mm}
\begin{equation} \label{equ:9}
\begin{aligned}
& s(\boldsymbol{a}, \boldsymbol{c}) = \cos(\arccos (s(\boldsymbol{a}, \boldsymbol{c}))) \\
&\geq \cos(\arccos (s(\boldsymbol{a}, \boldsymbol{b})) + \arccos (s(\boldsymbol{b}, \boldsymbol{c})))\\
& = s(\boldsymbol{a}, \boldsymbol{b})s(\boldsymbol{b}, \boldsymbol{c}) - 
\sin(\arccos (s(\boldsymbol{a}, \boldsymbol{b})))\sin(\arccos (s(\boldsymbol{b}, \boldsymbol{c}))).
\end{aligned}
\end{equation}

Recall for $\mathcal{L}_{i',1}$, malicious clients use the same clean model $\boldsymbol{\theta}$ as the benchmark to guide their backdoored models to generate correct embeddings for the reference image $\boldsymbol{x}_{target}$. Therefore, when $\boldsymbol{\theta_i}$ and $\boldsymbol{\theta_j}$ are backdoored, we can have $f(\boldsymbol{x}_{\text{target}}, \boldsymbol{\theta}_{i})\approx f(\boldsymbol{x}_{\text{target}}, \boldsymbol{\theta}_{j})$, $i\in\mathcal{M}$, $\forall j \in \mathcal{M}/i$.  

Without loss of generality, for $s(f(\boldsymbol{x}, \boldsymbol{\theta}_i), f(\boldsymbol{x}_{target}, \boldsymbol{\theta}_i)) = \sigma_i$ and $s(f(\boldsymbol{x}, \boldsymbol{\theta}_j), f(\boldsymbol{x}_{target}, \boldsymbol{\theta}_j)) = \sigma_j$, based on \eqref{equ:9} we have
\vspace{-2.5mm}
\begin{equation}
    s(f(\boldsymbol{x}, \boldsymbol{\theta}_i), f(\boldsymbol{x}, \boldsymbol{\theta}_j)) \geq \sigma_i\sigma_j - \sin(\arccos (\sigma_i))\sin(\arccos(\sigma_j)).
\end{equation}
As shown in Fig. \ref{fig:observation}, a backdoored model makes $\sigma_i$ and $\sigma_j$ larger and thus $\sin(\arccos (\sigma_i))$ and $\sin(\arccos(\sigma_j))$ become smaller. Therefore, the lower bound of $s(f(\boldsymbol{x}, \boldsymbol{\theta}_i), f(\boldsymbol{x}, \boldsymbol{\theta}_j))$ will become larger.
\end{proof}
\vspace{-3mm}
Therefore, backdoored models are more likely to have densely clustered (similar) embeddings.

\noindent
$\bullet$ \textbf{Embeddings from backdoored models are not outliers.} Backdoored models tend to generate embeddings clustered not too far from that generated by the global model, as there is a loss term (i.e., $L_{i',2}$) to constrain the backdoored model to maintain the accuracy of the primary task, which uses the initial clean global model as a benchmark.

\noindent
$\bullet$ \textbf{Embedings from backdoored models are over-stable regardless of the training data distribution.} For the majority of benign models, since they have no uniform objective function like $\mathcal{L}_{i',1}$ that pushes the model representation more similar to one target class, the amount of training data of each local client is limited, and the training involves lots of randomness (e.g., data augmentations), embedding generated by them exhibit divergence under both i.i.d and non-i.i.d settings. In contrast, the distribution of the training data would not change the stable nature of backdoor objectives, as we can observe that embeddings from malicious models clustered together and stay beside the embedding from the current global model under both i.i.d and non-i.i.d settings. Fig. \ref{fig: embedding space} demonstrates this observation while Tab \ref{tab3} and \ref{tab6} show the effectiveness of our method under both i.i.d and non-i.i.d settings. 

Finally, we can summarize that embeddings from backdoored models constantly congregate within the region of embeddings generated by benign models.

\subsection{Backdoor Mitigation for FSSL}
Recall the observations discussed above, for the same image, we find that the embeddings generated by backdoored models usually congregate in a certain place and intermingle with features extracted by benign models. Conversely, the distribution of embeddings generated by benign models is ruleless.
This means that embeddings generated by backdoored models are usually similar to those generated by all the other models, while there is randomness for benign models.
Based on this, we suggest detecting poisoned models by inspecting the embeddings generated by all uploaded local models and formally proposing the Embedding Inspector (EmInspector). In particular, EmInspector uses a set of inspection data to inspire the ``constancy'' of backdoored models and detect them based on the accumulated similarity values between output features from all the uploaded local models.

\begin{algorithm}[t]
\small
\caption{Embedding Inspector (EmInspector)}
\label{alg:RepGrard}
\SetKwData{Left}{left}\SetKwData{This}{this}\SetKwData{Up}{up} \SetKwFunction{Union}{Union}\SetKwFunction{FindCompress}{FindCompress}
\SetKwInOut{Input}{Input}\SetKwInOut{Output}{Output}
\Input{An honest server with an inspection dataset $\mathcal{D}_\text{ins}$ and the received local models $\boldsymbol{\theta}^{t}_{i}$, $i\in \mathcal{S}^{t}$.} 
\Output{Global model $\boldsymbol{\theta}^{t+1}$.}
\textbf{Initialization:} Malicious score $S_{i} \gets 0$, $i\in \mathcal{S}^{t}$ and $\overline{\mathcal{S}}^{t} \gets  \mathcal{S}^{t}$.\\
\For{$\boldsymbol{x}_{u}$ in $\mathcal{D}_\text{ins}$}{
    Calculate the accumulated similarity value $d_{i}$, 
    $i\in \mathcal{S}^{t}$ based on \eqref{eq:similarity}\\
    Obtain the decision boundary $\hat{d}$ based on \eqref{eq: db}\\
    \For{$i$ in  $\mathcal{S}^{t}$}{
        \If{$d_i(\boldsymbol{x}_{u})$ $\geq \hat{d}(\boldsymbol{x}_{u})$}{
                    $S_i = S_i + 1$
        }
        \Else{$S_i = S_i - 1$}
            }
        }       
\For{$i$ in  $\mathcal{S}^{t}$}{
    \If{$S_i > 0$}{
        Remove $i$ from $\overline{\mathcal{S}}^{t}$
    }
}
Obtain $\boldsymbol{\theta}^{t+1}$ by aggregating local models from the client set $\overline{\mathcal{S}}^{t}$
\end{algorithm}

Formally, the accumulated similarity of the output feature from the $i$-th model $\boldsymbol{\theta}_{i}$, $i\in\mathcal{N}$, can be defined as
\begin{equation}\label{eq:similarity}
\begin{aligned}
d_{i}(\boldsymbol{x}) &= \sum\limits_{j\in \mathcal{M}, j \neq i}s\left(f\left(\boldsymbol{x}, \boldsymbol{\theta}_{i}\right), f\left(\boldsymbol{x}, \boldsymbol{\tilde{\theta}}_{j}\right)\right) \\
&\quad + \sum\limits_{j\in \mathcal{N}/\mathcal{M}, j \neq i}s\left(f\left(\boldsymbol{x}, \boldsymbol{\theta}_{i}  \right), f\left(\boldsymbol{x}, \boldsymbol{\theta}_{j}\right)\right),
\end{aligned}
\end{equation}
where $\boldsymbol{x}$ is an image utilized for detecting backdoored models, referred to as an inspection image. Note that $\boldsymbol{\theta}_{i}$ can be uploaded by either benign or malicious clients and here we use $\boldsymbol{\theta}_i$ for simplicity.
With an inspection image, we calculate the average similarity value $d_{\text{mean}}$ and the median similarity value $d_{\text{median}}$ (we sort all values of \textit{d} and formulate a list $L_{d(\boldsymbol{x})}$ before calculating the median) for all local models as follows:
\begin{equation}\label{eq:avg}
{d_{\text{mean}}}(\boldsymbol{x}) = \frac{1}{N}\sum_{i \in \mathcal{N}} d_{i}(\boldsymbol{x}).
\end{equation}
\begin{equation} \label{eq: median}
		d_{\text{median}}(\boldsymbol{x}) =\left\{
		\begin{array}{cc}
			{L_{d(\boldsymbol{x})}}\left[\frac{N-1}{2}\right],   & if \,\,\,  {\text{\textit{N} is odd}},\\
                \\
			 \frac{{L_{d(\boldsymbol{x})}}\left[\frac{N}{2}-1\right] + {L_{d(\boldsymbol{x})}}\left[\frac{N}{2}\right]}{2},   & if \,\,\,  {\text{\textit{N} is even}}.\\
		\end{array} 
        \right.
\end{equation}
As we assumed, benign clients comprise the majority among all local clients in the FSSL system, therefore EmInspector updates the malicious score by using the bigger value among $d_{\text{mean}}$ and $d_{\text{median}}$ as the decision boundary $\hat{d}(\boldsymbol{x})$:
\begin{equation} \label{eq: db}
    \hat{d}(\boldsymbol{x}) = \max(d_{\text{mean}}(\boldsymbol{x}), d_{\text{median}}(\boldsymbol{x})).
\end{equation}
Compared to directly using Eq. \ref{eq:avg} as the decision boundary, Eq. \ref{eq: db} decreases the possibility of misclassifying benign clients as malicious when there exist very few or no malicious clients.
That is Eq. \ref{eq: db} ensures that EmInspector regards less than 50\% of uploaded local models as malicious with each inspection data fed. Furthermore, in Sec. \ref{eval: defense}, we conduct an ablation study to show that Eq. \ref{eq: db} decreases the false positive rate from 60\% to 35\% on CIFAR10 when there is no malicious client. 
Formally, EmInspector then updates the malicious score by comparing the similarity of each local model with $\hat{d}(\boldsymbol{x})$:
\begin{equation}
		S_i =\left\{
		\begin{array}{cc}
			S_i + 1,   & if \,\,\,  {d_{i}(\boldsymbol{x}) \geq  \hat{d}(\boldsymbol{x})},\\
			S_i - 1,   & if \,\,\,  {d_{i}(\boldsymbol{x}) <  \hat{d}(\boldsymbol{x})}.\\
		\end{array} 
        \right.
\end{equation}

Assume the server has an inspection dataset $\mathcal{D}_{\text{ins}}$\footnote{This is practical since the inspection data is not required to conform to a specific data distribution or be labeled. The defender can collect from the web or use the publicly available dataset.}, EmInspector will repeat this update process for $\vert \mathcal{D}_{\text{ins}} \vert$ times.
Finally, the server will regard client $i$ as malicious if its final malicious score $S_i > 0$, and as benign otherwise. The server will remove all models regarded as backdoored totally and aggregate the remaining. 
The detailed steps of the proposed EmInspector algorithm are summarized in Algorithm \ref{alg:RepGrard}.

\section{Experimental Setup} \label{sec: exp setup}
\subsection{Datasets and Models}
Our evaluation primarily uses three classic datasets, i.e., CIFAR10 \cite{krizhevsky2009learning}, STL10 \cite{coates2011analysis}, and CIFAR100 \cite{krizhevsky2010cifar}. Besides, to evaluate backdoor attacks under the pre-training and downstream datasets being distinct, we also include GTSRB \cite{stallkamp2012man} as an additional dataset. The detailed descriptions of these datasets are provided in Appendix \ref{sec:appendix for setup}.

For most experiments, we employ SimCLR to train a widely-used ResNet18 \cite{he2016deep} as the model for jointly learning a global encoder. 
We also evaluate Vgg19\_bn \cite{simonyan2014very}, ResNet34 \cite{he2016deep} and ResNet50 \cite{he2016deep} in Sec. \ref{sec: experimental results}.

\subsection{FSSL Setting}
We consider an FSSL system consisting of $25$ clients, with each client having $2,000$ unlabeled images for local training.
We adopt both i.i.d. and non-i.i.d. data settings. For the non-i.i.d. setting, we randomly assign two classes of CIFAR10 or four classes of CIFAR100 to each client. 
For the i.i.d setting, we randomly sample 2,000 images from the training set of CIFAR10/CIFAR100 for each client.  
For STL10, we randomly sample $2,000$ images for each client from its 100,000 unlabeled images throughout our experiments.
Following previous works \cite{Xie2020DBA, haoyang20233dfed}, without mentioning otherwise, we adopt the non-i.i.d setting for our experiments.
In Sec. \ref{sec: experimental results}, we demonstrate the vulnerability of FSSL to backdoor attacks and the effectiveness of EmInspector under an i.i.d data setting over CIFAR10 and CIFAR100.

Throughout the training phase, the server randomly selects $10$ clients from the overall $25$ local clients for joint training and aggregation in each round, where each client trains its local model for $10$ epochs using the Adam optimizer \cite{kingma2014adam} with a learning rate of $0.001$.

\begin{figure*}[t]
\centering
\includegraphics[width=0.85\textwidth]{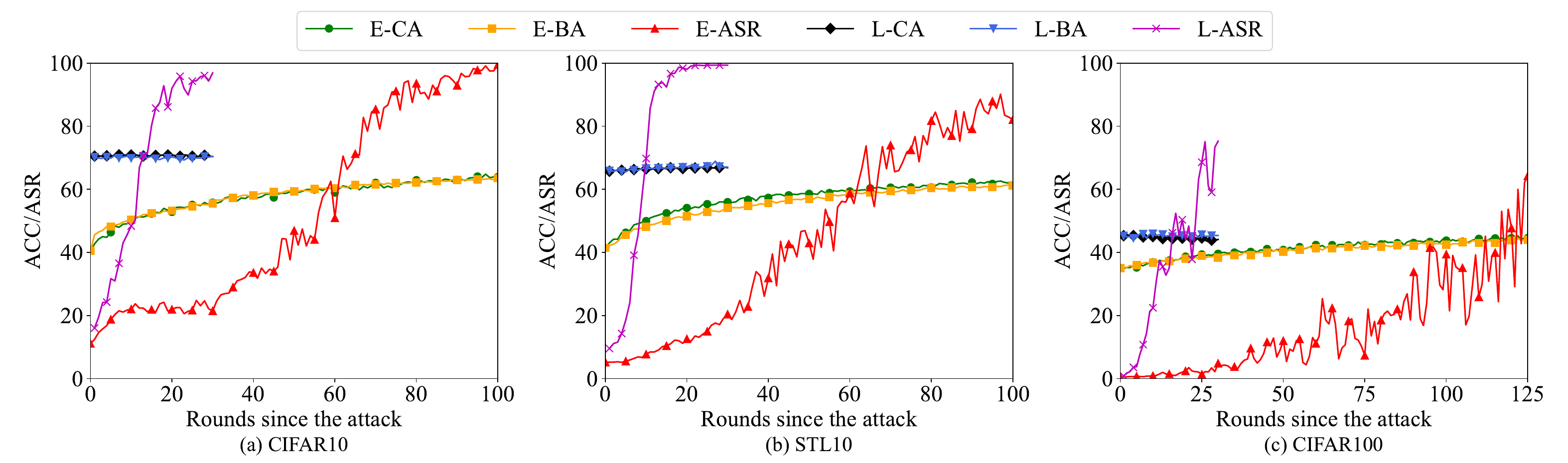}
\caption{Attacking from the very beginning v.s. attacking when the global model is close to convergence. CA represents the ACC of the clean model under a no-attack setting. BA represents the ACC of the backdoored model. ASR represents the attack success rate. The prefix ``E'' means the former ``attack'' early scenario and the prefix ``L'' means the latter ``attack late'' scenario.}
\label{attack from beginning}
\end{figure*}

\begin{figure*}[t]
\centering
\includegraphics[width=0.85\textwidth]{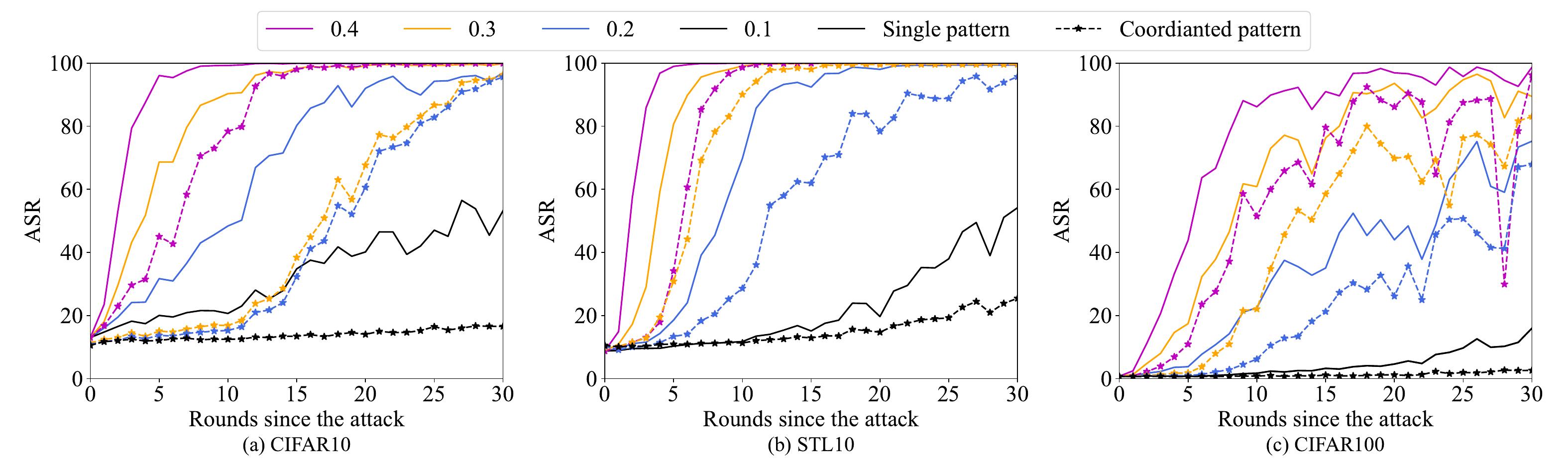}
\caption{Backdoor attacks on FSSL with different proportions of malicious clients (10\% - 40\%) and collusion methods.}
\label{attack_results}
\end{figure*}

\subsection{Attack Setting}
We evaluate the impact of the proportion (prop.) of malicious clients (MC), ranging from 10\% to 40\%, as well as the collusion methods, single or coordinated pattern, on the discussed backdoor attack.
Referring to \cite{jia2022badencoder}, during backdoor training, we set both $\lambda_1$ and $\lambda_2$ in \eqref{eq:trade off} to $1$ and use the SGD optimizer with a learning rate of $0.05$ to launch attacks for 30 rounds. 
Without other mentioned, we follow a practical setting similar to the previous work \cite{Xie2020DBA}, in which the attacker begins to attack when the global model is close to convergence. 
We choose ``truck'', ``airplane'', ``camel'' and ``priority road sign'' as our target class for CIFAR10, STL10, CIFAR100, and GTSRB, respectively. In particular, malicious clients will embed triggers into all images in their local training datasets to maximize the efficiency of the attack. 
The triggers and the reference images of the target classes we use are illustrated in Fig. \ref{fig:triggers} and \ref{fig:reference images} in Appendix \ref{sec:appendix for setup}, respectively.
Moreover, we freeze the parameters of batch normalization (BN) layers when launching backdoor attacks for efficiency and persistence (refer to Fig. \ref{fig:freeze bn} in Appendix. \ref{appendix:experiment results}).

\subsection{Defense Setting}
In practice, malicious clients may train local models for $\eta$ epochs standardly like benign clients first, and then launch the backdoor attack for the remaining $(10-\eta)$ epochs, which can make the attack more difficult to detect. To simulate a stealthier attack, we consider an FSSL system consisting of 20\% malicious clients and set $\eta$ to $1$ by default when evaluating defense methods, which means the parameters of BN layers of backdoored models are distinct.

\noindent
\textbf{Baselines.} We compare our proposed EmInspector with seven baselines as follows.
\\$\bullet$ \textbf{Krum} \cite{blanchard2017machine}: chooses one model that of the smallest Euclidean distances with others.
\\$\bullet$ \textbf{Trimmed-Mean} \cite{Dong2018Byzantine}: update each parameter based on the trimmed mean of uploaded models in a coordinate-wise way.
\\$\bullet$ \textbf{FLTrust} \cite{Cao2021FLTrust}: the server holds a trusted model to exclude models that diverge too much from it. 
\\$\bullet$ \textbf{Foolsgold} \cite{Fung2020}: punishes clients that have historical updates with high pairwise cosine similarities by a low learning rate.
\\$\bullet$ \textbf{FLAME} \cite{nguyen2022flame}: bounds the added noise of differential privacy to mitigate the negative impact on primary tasks. 
\\$\bullet$ \textbf{RFLBAT} \cite{wang2022rflbat}: applies Principal Component Analysis (PCA) to separate malicious models from benign models in a low dimensional space. 
\\$\bullet$ \textbf{FLARE} \cite{wang2022flare}: uses the Maximum Mean Discrepancy (MMD) to detect outliers as malicious in the latent representations of uploaded local models.

For EmInspector, we randomly sample $100$ images from the training set to form the inspection dataset, without requiring labels. Additionally, in Sec. \ref{eval: defense}, we present ablation results to show the effects of varying the number of inspection images.

\subsection{Evaluation Metrics}
We adopt two metrics to evaluate the trained global model as follows:
\\\textbf{Primary task accuracy (ACC):}
ACC indicates the classification accuracy of the global SSL model for clean images. 
\\\textbf{Attack success rate (ASR):}
ASR indicates the ratio of triggered images that are misclassified as the attacker's desired target classes.

In addition, we also need to ensure the convergence speed and training performance of the primary task when backdoor attacks occur during training, for which we use a KNN monitor \cite{wu2018unsupervised} to test the accuracy of the primary task and the success rate of the backdoor attack.

\section{Experimental Results} \label{sec: experimental results}
In this section, we conduct experiments to answer three vital questions. In Sec. \ref{eval: attack}, we answer: $\textbf{Q1}$: Is the discussed backdoor attack paradigm effective against FSSL? If so, what factors impact its effectiveness and efficiency? In Sec. \ref{eval: defense}, we answer: $\textbf{Q2}$: Can existing defense mechanisms mitigate backdoor attacks on FSSL? $\textbf{Q3}$: Is EmInspector capable of mitigating backdoor attacks on FSSL? If so, what factors impact its performance?   

\begin{table}[t]
\centering
\caption{Backdoor attacks on FSSL with different proportions of malicious clients and collusion methods.}
\resizebox{\linewidth}{!}{\begin{tabular}{@{}c|c|cc|cc|cc@{}}
\toprule[1.2pt]
\multirow{2}{*}{\begin{tabular}[c]{@{}c@{}}Collusion\\ Method\end{tabular}} &
  \multirow{2}{*}{\begin{tabular}[c]{@{}c@{}}Prop.\\ of MCs\end{tabular}} &
  \multicolumn{2}{c|}{CIFAR10} &
  \multicolumn{2}{c|}{STL10} &
  \multicolumn{2}{c}{CIFAR100} \\ \cmidrule(l){3-8} 
 &
   &
  ACC &
  ASR &
  ACC &
  ASR &
  ACC &
  ASR \\ \midrule[0.75pt]
None &
  0\% &
  \multicolumn{1}{c|}{70.52} &
  13.03 &
  \multicolumn{1}{c|}{66.00} &
  8.76 &
  \multicolumn{1}{c|}{45.16} &
  0.75 \\ \midrule
\multirow{4}{*}{\begin{tabular}[c]{@{}c@{}}Single\\ Pattern\end{tabular}} &
  10\% &
  \multicolumn{1}{c|}{70.40} &
  56.49 &
  \multicolumn{1}{c|}{67.06} &
  54.16 &
  \multicolumn{1}{c|}{44.06} &
  15.97 \\
 &
  20\% &
  \multicolumn{1}{c|}{70.35} &
  96.89 &
  \multicolumn{1}{c|}{66.84} &
  99.53 &
  \multicolumn{1}{c|}{44.11} &
  75.25 \\
 &
  30\% &
  \multicolumn{1}{c|}{70.23} &
  99.80 &
  \multicolumn{1}{c|}{66.51} &
  99.98 &
  \multicolumn{1}{c|}{44.14} &
  96.48 \\
 &
  40\% &
  \multicolumn{1}{c|}{69.91} &
  99.98 &
  \multicolumn{1}{c|}{64.40} &
  100 &
  \multicolumn{1}{c|}{44.00} &
  98.70 \\ \midrule
\multirow{4}{*}{\begin{tabular}[c]{@{}c@{}}Coordinated\\ Pattern\end{tabular}} &
  10\% &
  \multicolumn{1}{c|}{71.22} &
  16.59 &
  \multicolumn{1}{c|}{67.20} &
  25.45 &
  \multicolumn{1}{c|}{44.41} &
  2.16 \\
 &
  20\% &
  \multicolumn{1}{c|}{70.54} &
  95.59 &
  \multicolumn{1}{c|}{67.23} &
  95.69 &
  \multicolumn{1}{c|}{44.11} &
  67.92 \\
 &
  30\% &
  \multicolumn{1}{c|}{70.49} &
  96.32 &
  \multicolumn{1}{c|}{66.65} &
  99.58 &
  \multicolumn{1}{c|}{44.20} &
  83.05 \\
 &
  40\% &
  \multicolumn{1}{c|}{70.10} &
  99.93 &
  \multicolumn{1}{c|}{66.95} &
  99.95 &
  \multicolumn{1}{c|}{44.12} &
  96.22 \\ \bottomrule[1.2pt]
\end{tabular}}
\label{tab1}
\end{table}

\subsection{AQ1 (Answer Q1): Attack Results} \label{eval: attack}

\noindent
\textbf{Attacking early v.s. Attacking later.} To verify the impact of the attack timing, we consider two scenarios: attacking from the beginning vs.. attacking when the global model is close to convergence.
We employ a single pattern attack with 20\% malicious clients for evaluation.
For the former scenario, we initiate the attack at round 5 for CIFAR10 and STL10, and at round 35 for CIFAR100, respectively. For the latter, we will attack when the global model is nearing convergence. To simplify our description, we refer to the former scenario as ``attack early'' and the latter as ``attack late''.
In Fig. \ref{attack from beginning}, we illustrate the training process after attacks begin. We observe that in both scenarios, both the ACC and ASR increase along with the training process, while the ASR converges much faster if we attack later. To our surprise, attacking at an early stage of network training does not seem to negatively impact the primary task accuracy, contrary to the understanding of previous work \cite{Xie2020DBA}. This counterintuitive result indicates that the investigated backdoor attack can be launched at any round without hurting the performance of primary tasks, thereby highlighting the severity of backdoor attacks on FSSL. 

\noindent
\textbf{Comparison of different prop. of MCs.} The top row of Tab. \ref{tab1} displays the ACC of the initial clean global model, along with the initial ASR triggered by the global trigger pattern. This serves as a baseline for evaluating the efficacy of backdoor attacks on FSSL. The subsequent rows exhibit ACCs and ASRs of backdoored global models that are trained with various proportions of malicious clients and two different collusion methods.
Notably, both single and coordinated pattern attacks achieve ASRs of over 95\% once the proportion of malicious clients reaches 20\% for CIFAR10 and STL10 datasets. With a lower proportion of 10\%, such attacks can still achieve ASRs of over 50\%.
Despite being a more intricate dataset, CIFAR100 is also vulnerable to backdoor attacks, with 20\% of malicious clients achieving an ASR of approximately 70\%.
Moreover, all attacks have a negligible impact on the primary task accuracy, making it difficult to notice the existence of backdoors. We also depict the training process in Fig. \ref{attack_results}, revealing that as the proportion of malicious clients increases, the ASR converges faster.

\noindent
\textbf{Comparison of two collusion methods.} According to Fig. \ref{attack_results}, it is evident that ASRs in single pattern attacks converge faster compared to in coordinate pattern attacks when the proportion of malicious clients is identical. 
This can be attributed to the fact that all malicious clients in a single pattern attack employ the same trigger to backdoor the global model, thereby enhancing the effect of this trigger, let alone the trigger used in a single pattern attack is larger, being easier for the model to perceive. 
Although coordinated pattern attacks are less efficient, they may bring more challenges to the design of defense mechanisms as each malicious client uses a unique trigger to execute the attack, making the changes of local models of more variety.  

\noindent
\textbf{The fluctuation of ASRs on CIFAR100.} In Fig. \ref{attack_results} (c), it is worth noting that ASRs exhibit considerable fluctuations when they approach a relatively high level on CIFAR100. We interpret this phenomenon as being caused by the extensive number of classes in CIFAR100, some of which may share common features, resulting in minor differences between them. Consequently, the attacker might overfine-tune the model with our preset learning rate, leading to a rapid decline in the ASR.

\begin{table}[t]
\centering
\caption{The performance of backdoor attacks on FSSL under pre-training and downstream datasets are distinct.}
\resizebox{\linewidth}{!}{\begin{tabular}{@{}c|c|cc|cc@{}}
\toprule[0.95pt]
\multirow{2}{*}{\begin{tabular}[c]{@{}c@{}}Training\\ Dataset\end{tabular}} &
  \multirow{2}{*}{\begin{tabular}[c]{@{}c@{}}Downstream\\ Dataset\end{tabular}} &
  \multicolumn{2}{c|}{Clean} &
  \multicolumn{2}{c}{Backdoored} \\ \cmidrule(l){3-6} 
                          &         & ACC                        & ASR   & ACC                        & ASR   \\ \midrule[0.6pt]
\multirow{2}{*}{CIFAR10}  & STL10   & \multicolumn{1}{c|}{66.95} & 10.70 & \multicolumn{1}{c|}{68.46} & 99.37 \\
                          & GTSRB   & \multicolumn{1}{c|}{75.53} & 4.79  & \multicolumn{1}{c|}{76.18} & 98.12 \\ \midrule
\multirow{2}{*}{STL10}    & CIFAR10 & \multicolumn{1}{c|}{80.81} & 8.92  & \multicolumn{1}{c|}{81.14} & 96.87 \\
                          & GTSRB   & \multicolumn{1}{c|}{69.07} & 6.36  & \multicolumn{1}{c|}{68.81} & 81.66 \\ \midrule
\multirow{2}{*}{CIFAR100} & CIFAR10 & \multicolumn{1}{c|}{80.49} & 11.72 & \multicolumn{1}{c|}{80.71} & 99.15 \\
                          & GTSRB   & \multicolumn{1}{c|}{76.73} & 5.73  & \multicolumn{1}{c|}{76.84} & 99.97 \\ \bottomrule[0.95pt]
\end{tabular}}
\label{tab2}
\end{table}

\begin{figure}[t]
\centering
\includegraphics[width=\linewidth]{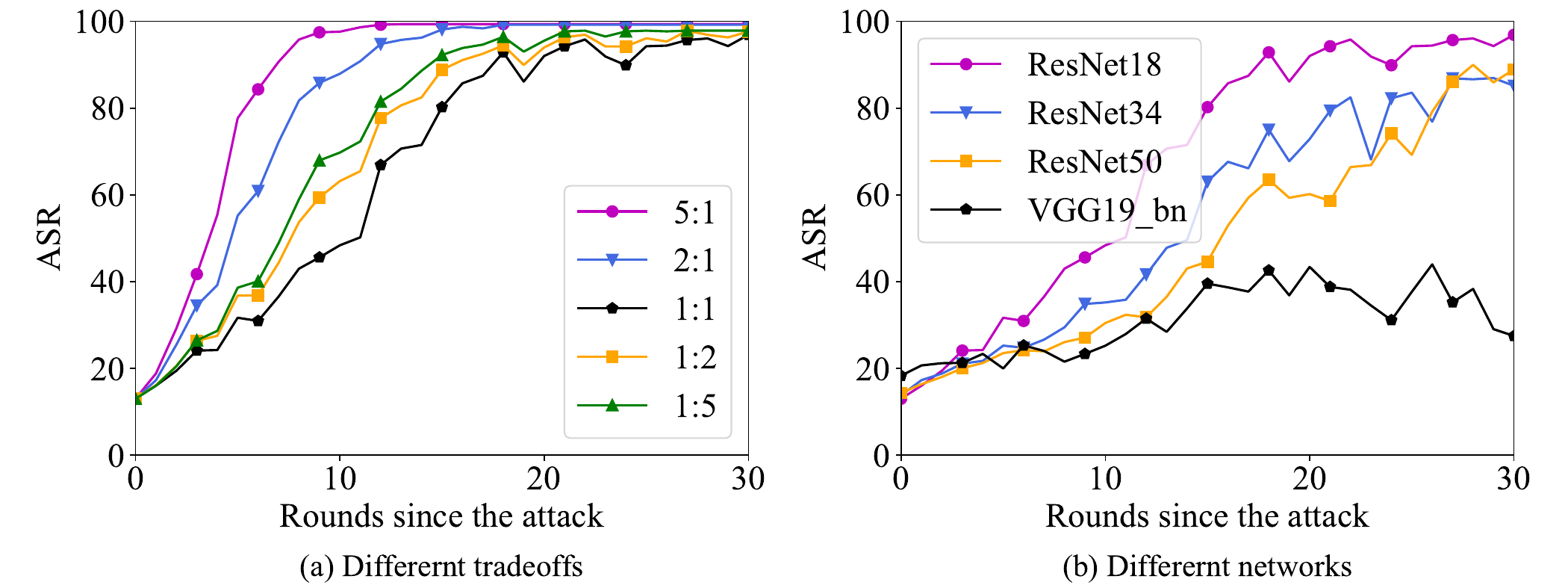}
\caption{Attack performance with (a) different attack degrees and (b) on different network architectures on CIFAR10.}
\label{fig:different networks}
\end{figure}

\noindent
\textbf{Comparison of different attack degrees ($\lambda$)}
To verify the impact of the attack degree, we conduct an ablation experiment with five different tradeoffs between the values of $\lambda_1$ and $\lambda_2$, i.e., 5 : 1, 2 : 1, 1 : 1, 1 : 2, and 1 : 5, where a larger value of $\lambda_1$ : $\lambda_2$ indicates a more aggressive attack. 
Our results are presented in Fig. \ref{fig:different networks} (a). We can observe that increasing the value of $\lambda_1$ : $\lambda_2$ leads to faster convergence of the ASR, which can be expected. However, to our surprise, when $\lambda_1$ is smaller than $\lambda_2$, the ASR still converges faster than when they are equal. We suspect the reason is that a lower value helps the local model maintain a better learning representation ability, which helps to perceive triggers and identify reference data more accurately.

\noindent 
\textbf{Attacking on different neural networks.} 
\label{subsec: Attacking different networks}
To evaluate the efficacy of backdoor attacks on FSSL across different network architectures, we test a single pattern attack with 20\% malicious clients on CIFAR10 with four distinct global model architectures, i.e. ResNet18, ResNet34, ResNet50, and VGG19\_bn.
Note that before launching an attack, the image encoders of different networks have been pre-trained for identical rounds. Our results are presented in Fig. \ref{fig:different networks} (b), proving that the investigated backdoor attacks are effective across different network architectures. 
Moreover, we observe that when the pre-trained global models are of similar ACCs (see Tab. \ref{tab:RepGuard across different networks}), the ASR on ResNet18 converges faster than on ResNet34, which in turn, converges faster than on ResNet50. This suggests that attacks on more complex networks exhibit a slower convergence speed. We also notice that the ASR on VGG19\_bn is significantly lower than on ResNets. Our interpretation for this is that the residual connection in ResNet helps to build a shortcut between the target class with the backdoor trigger, as also studied in \cite{yang2023backdoor}.

\noindent
\textbf{Attacking under pre-training and downstream datasets different.}
Tab. \ref{tab2} shows that when the pre-training and downstream datasets are different, backdoor attacks still retain a high ASR in the downstream classifier. For example, with CIFAR10 and GTSRB as the pre-training dataset and downstream dataset, the backdoored model and the clean model achieve ASRs of 98.12\% and 4.79\%, respectively. This large gap between the two ASRs indicates a highly effective backdoor attack. Moreover, the accuracy of the backdoored model on clean inputs is not affected, indicating the backdoor attack of high utility (i.e., the attack is stealth).

\begin{table}[t]
\centering\caption{Defense performance of EmInspector and existing defense methods under a setting of 20\% malicious clients. Results of EmInspector are marked in bold.}
\resizebox{\linewidth}{!}{\begin{tabular}{@{}c|c|cc|cc|cc@{}}
\toprule[1.2pt]
\multirow{2}{*}{\begin{tabular}[c]{@{}c@{}}Collusion\\ Method\end{tabular}} &
  \multirow{2}{*}{\begin{tabular}[c]{@{}c@{}}Defense\\ Method\end{tabular}} &
  \multicolumn{2}{c|}{CIFAR10} &
  \multicolumn{2}{c|}{STL10} &
  \multicolumn{2}{c}{CIFAR100} \\ \cmidrule(l){3-8} 
                  &  & ACC & ASR & ACC & ASR & ACC & ASR \\ \midrule[0.75pt]
\multirow{7}{*}{\begin{tabular}[c]{@{}c@{}}Single\\ Pattern\end{tabular}} & Krum~\cite{blanchard2017machine} &  \multicolumn{1}{c|}{69.37} & 100 &  \multicolumn{1}{c|}{65.51} & 100 & \multicolumn{1}{c|}{42.78}    &  99.97   \\
                  & Trimmed-Mean ~\cite{Dong2018Byzantine} &  \multicolumn{1}{c|}{70.10}   &  99.47   &  \multicolumn{1}{c|}{66.48}   &  99.31   &  \multicolumn{1}{c|}{44.17}   &  97.68  \\
                  & FLTrust~\cite{Cao2021FLTrust} &  \multicolumn{1}{c|}{70.57} &  99.51 &  \multicolumn{1}{c|}{65.91}  &  99.83   &  \multicolumn{1}{c|}{44.79}    &  99.33   \\
                  & Foolsgold~\cite{Fung2020} &  \multicolumn{1}{c|}{69.96}   &  100  &  \multicolumn{1}{c|}{65.87} & 100  &  \multicolumn{1}{c|}{44.15}    &  86.63   \\
                  & FLAME~\cite{nguyen2022flame} &  \multicolumn{1}{c|}{70.04}   &   99.40  &  \multicolumn{1}{c|}{66.18} & 92.94   & \multicolumn{1}{c|}{44.50}   &  94.30   \\
                  & RFLBAT~\cite{wang2022rflbat} &  \multicolumn{1}{c|}{69.21}   &  99.80   &  \multicolumn{1}{c|}{65.71} & \multicolumn{1}{c|}{99.18}     & \multicolumn{1}{c|}{44.33}    &  97.34   \\
                  & FLARE~\cite{wang2022flare} &  \multicolumn{1}{c|}{69.44}   &  100   &  \multicolumn{1}{c|}{65.65}   &  99.81   &   \multicolumn{1}{c|}{44.79} &   100  \\
                  & \textbf{EmInspector} & \multicolumn{1}{c|}{\textbf{69.52}}    &  \textbf{17.71}   &   \multicolumn{1}{c|}{\textbf{66.80}}  &  \textbf{8.78}   &  \multicolumn{1}{c|}{\textbf{43.62}}   &  \textbf{0.21}   \\ \midrule
\multirow{7}{*}{\begin{tabular}[c]{@{}c@{}}Coordinated \\ Pattern\end{tabular}} & Krum~\cite{blanchard2017machine} & \multicolumn{1}{c|}{69.62}    &  99.43  &   \multicolumn{1}{c|}{65.59}  & 99.80    &  \multicolumn{1}{c|}{42.36}   &  92.08   \\
                  & Trimmed-Mean ~\cite{Dong2018Byzantine} &  \multicolumn{1}{c|}{70.22}   &   97.33  &  \multicolumn{1}{c|}{66.68}   &  98.40   &   \multicolumn{1}{c|}{44.13}  &  86.79   \\
                  & FLTrust~\cite{Cao2021FLTrust} &   \multicolumn{1}{c|}{70.61}  & 98.33 & \multicolumn{1}{c|}{65.95}    &  99.95   &   \multicolumn{1}{c|}{44.42}  &   99.79  \\
                  & Foolsgold~\cite{Fung2020} &  \multicolumn{1}{c|}{70.63}   &  100   &   \multicolumn{1}{c|}{65.40}  &  100   &  \multicolumn{1}{c|}{43.22}   &  73.10   \\
                  & FLAME~\cite{nguyen2022flame} &   \multicolumn{1}{c|}{68.92}  &  98.67   &  \multicolumn{1}{c|}{65.94}   &   96.30  &   \multicolumn{1}{c|}{44.02}  &  87.57   \\
                  & RFLBAT~\cite{wang2022rflbat} &  \multicolumn{1}{c|}{69.70}   &  99.93   &  \multicolumn{1}{c|}{66.17}   &   99.93  &   \multicolumn{1}{c|}{44.51} &  99.69   \\
                   & FLARE~\cite{wang2022flare} &  \multicolumn{1}{c|}{69.37}   &  100   &  \multicolumn{1}{c|}{66.03}   &  99.27   &   \multicolumn{1}{c|}{44.12} &   99.83  \\
                  & \textbf{EmInspector} &  \multicolumn{1}{c|}{\textbf{70.92}} & \textbf{13.28}  & \multicolumn{1}{c|}{\textbf{67.16}}    & \textbf{10.25}   &  \multicolumn{1}{c|}{\textbf{44.47}}   &  \textbf{0.42}   \\ \bottomrule[1.2pt]
\end{tabular}}
\label{tab3}
\end{table}

\begin{table}[t] 
\centering
\caption{The performance (ACC/ASR) of EmInspector under i.i.d data setting over CIFAR10 and CIFAR100.}
\label{tab: iid exp}
\begin{tabular}{@{}c|c|c|c@{}}
\toprule[1.2pt]
Dataset & \begin{tabular}[c]{@{}c@{}}Prop. of  MCs\end{tabular} & \begin{tabular}[c]{@{}c@{}}No Defense\end{tabular} & EmInspector \\ \midrule[0.75pt]
\multirow{2}{*}{CIFAR10}  & 20\% & 78.65 / 91.42 & 79.78 / 12.09 \\
                          & 30\% & 77.96 / 98.10 & 79.61 / 12.30 \\ \midrule
\multirow{2}{*}{CIFAR100} & 20\% & 51.24 / 83.56 & 51.42 / 0.94 \\
                          & 30\% & 50.71 / 95.81 & 51.23 / 0.82 \\ \bottomrule[1.2pt]
\end{tabular}
\label{tab6}
\end{table}

\begin{table}[t]
\centering
\caption{The performance (ACC/ASR) of EmInspector under different attack degrees on CIFAR10.}
\label{tab:different lamda}
\resizebox{\linewidth}{!}{\begin{tabular}{@{}c|c|c|c|c|c@{}}
\toprule[1.2pt]
Defense &  5 : 1 & 2 : 1 & 1 : 1 & 1 : 2 & 1 : 5  \\ \midrule[0.75pt]
None & 70.60/99.36 & 70.84/99.21 & 70.35/96.89 & 70.23/97.65 & 70.77/97.85 \\ \midrule
EmInspector &  71.05/13.79 & 69.81/14.33 & 69.52/17.71 & 70.07/13.63 & 71.33/14.24 \\ \bottomrule[1.2pt]
\end{tabular}}
\label{tab4}
\end{table}

\begin{table}[t]
\centering
\caption{The performance (ACC/ASR) of EmInspector across different network architectures on CIFAR10.}
\label{tab:RepGuard across different networks}
\resizebox{\linewidth}{!}{\begin{tabular}{@{}c|c|c|c|c@{}}
\toprule[1.2pt]
Defense & ResNet18 & ResNet34 & ResNet50 & VGG19\_bn  \\ \midrule[0.75pt]
None & 70.35/96.89 & 67.67/85.14 & 68.73/88.93 & 46.85/43.99 \\ \midrule
EmInspector & 69.52/17.71 & 67.13/14.28 & 68.61/16.02 & 47.64/13.58 \\ \bottomrule[1.2pt]
\end{tabular}}
\label{tab5}
\end{table}

\subsection{Defense Results} \label{eval: defense}
\textbf{AQ2\&3: Effectiveness of EmInspector.}
Tab. \ref{tab3} shows the performance of our EmInspector against seven existing defense methods under two different collusion methods and the non-i.i.d setting. From the results, it is evident that despite the deployment of the existing defenses, the global model still is infiltrated by malicious clients, exhibiting backdoor behaviors. This observation leads to the conclusion that existing defense strategies are inadequate in mitigating backdoor attacks on FSSL. On the one hand, the failures of statistical-based methods (i.e., the former six baselines in Tab. \ref{tab3}) indicate that the parameters of backdoored models have minimal deviations from those of benign models. On the other hand, though FLARE also detects backdoored models by inspecting local models' embedding space, it is based on an observation that is untenable in FSSL, which is exactly the opposite of ours. This reveals that embeddings from backdoored models are not outliers among those from benign models. In contrast, our EmInspector successfully maintains ASRs at a low level without obviously degrading the primary task accuracy across all datasets. Specifically, EmInspector ensures the increase of the ASR of no more than 5\% on CIFAR10, no more than 2\% on STL10, and maintains the ASR on CIFAR100 at a level of nearly 0\%. 
Moreover, the degradation of the primary task accuracy is less than 1\% on CIFAR10 and less than 1.54\% on CIFAR100, whereas, on STL10, the primary task accuracy increases by about 1\%. Our findings demonstrate that EmInspector outperforms existing defense methods by a significant margin in mitigating backdoor attacks on FSSL. 

In addition, it is common for multiple groups of clients to have very similar local data distributions in the real world. Therefore, we conduct some experiments under an i.i.d data setting across CIFAR10 and CIFAR100 to further demonstrate the vulnerability of FSSL to backdoor attacks and the effectiveness of EmInspector. For attacks, we adopt the single pattern attack of 1:1 attack degree with two different proportions of malicious clients, namely 20\% and 30\%. Results are presented in Tab. \ref{tab6}, where we can see that the investigated backdoor attack is also of high efficacy, achieving ASRs of over 80\% on both datasets. Furthermore, the gap of ASRs between the FSSL system deployed with EmInspector and with FedAvg (e.g., on CIFAR10 with 20\% malicious clients, ASR of 12.09 vs. 91.42) indicates that EmInspector remains highly effective in detecting malicious clients even though under the i.i.d data settings. 

\begin{figure}[t]
\centering
\includegraphics[width=\linewidth]{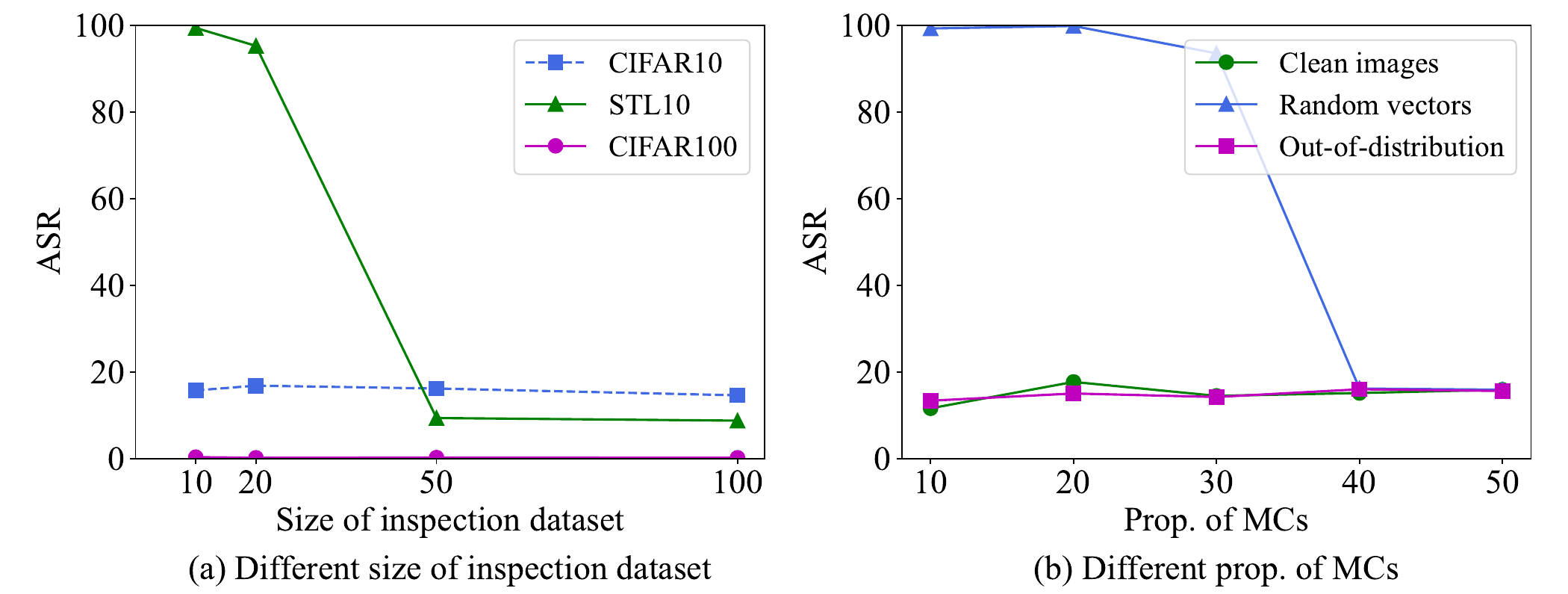}
\caption{EmInspector (a) with different numbers of inspection images and (b) under different proportions of malicious clients, using clean images, out-of-distribution data, or randomly generated vectors as the inspection data.}
\label{impact of the number of detection images and prop of mc}
\end{figure}


\noindent
\textbf{AQ3: Defense performance under different attack degrees.}
We evaluate our EmInspector under attacks with five different tradeoffs between $\lambda_1$ and $\lambda_2$, using the same setting as we evaluate the backdoor attacks. Tab. \ref{tab4} shows that EmInspector stays effective across various attack degrees, retaining the ASRs of attacks of various attack degrees at a low level.
Moreover, the primary task accuracy is not affected.

\noindent
\textbf{AQ3: Defense performance on different networks.}
\label{sec:fldiscrimitaor across different networks} 
We also evaluate our EmInspector on four distinct network architectures, using the same setting as we evaluate the backdoor attack.
Tab. \ref{tab5} shows that EmInspector retains ASRs on all networks at a low level, which demonstrates its effectiveness across various network architectures. Moreover, the primary task accuracy is not affected.

\noindent
\textbf{AQ3: Defense performance with different numbers of inspection images.}
EmInspector does not require the inspection data to conform to a specific sample distribution. Thus we characterize the inspection dataset by its size.
Fig. \ref{impact of the number of detection images and prop of mc} (a) shows ASRs under a single pattern attack with 20\% malicious clients when the size of the inspection dataset increases from 10 to 100. Our results indicate that using only ten inspection images is sufficient for EmInspector to effectively mitigate backdoors on CIFAR10 and CIFAR100. For STL10, though a larger inspection dataset is required, 50 is enough. These findings demonstrate that EmInspector with very few inspection images is effective in detecting malicious clients.

\noindent
\textbf{AQ3: Defense performance under different proportions of MCs.}
A test is done on CIFAR10 to evaluate the effectiveness of EmInspector under adversary settings of various proportions of malicious clients, ranging from 10\% to 50\%. As depicted in Fig. \ref{impact of the number of detection images and prop of mc} (b), EmInspector is robust against various ratios of malicious clients. Moreover, we use 100 images sampled from the GTSRB dataset to investigate the effectiveness of using out-of-distribution data as the inspection data. The purple curve in Fig. \ref{impact of the number of detection images and prop of mc} (b) indicates that even though the inspection data is out-of-distribution with the training data, EmInspector still works effectively.   
In addition, we substitute the 100 inspection images with 100 randomly generated vectors of the same dimension with an inspection image and discover that EmInspector can still perform effectively when the proportion of malicious clients reaches 40\%, as demonstrated by the blue curve in Fig. \ref{impact of the number of detection images and prop of mc} (b). However, EmInspector performs poorly when the proportion of malicious clients is below this threshold and it is more prone to make misclassifications than using clean images as inspection data. This approach can serve as a remedy when the server lacks access to clean images but has an approximation of the ratio of malicious clients.

\begin{table}[t]
\centering
\caption{The performance of EmInspector on identification accuracy with 100 inspection images on CIFAR10.}
\label{tab: identification acc}
\resizebox{\linewidth}{!}{\begin{tabular}{@{}c|c|c|c|c|c|c|c@{}}
\toprule[1.2pt]
Prop. of MCs &
  \begin{tabular}[c]{@{}c@{}}Attack\\ Degree\end{tabular} &
  FPR &
  TPR &
  B-MS &
  M-MS &
  \begin{tabular}[c]{@{}c@{}}ACC\\ (Avg)\end{tabular} &
  \begin{tabular}[c]{@{}c@{}}ACC\\ (Ours)\end{tabular} \\ \midrule[0.75pt]
0\%                   & \multirow{2}{*}{--}     & 35\% & -- & -5.22 & -- & 71.06 & 70.64 \\
0\% (Eq. \ref{eq:avg})       &                       & 60\% & -- & 3.86 & -- & 71.06 & 69.88 \\ \midrule
10\%                  & \multirow{4}{*}{1 : 1} & 20\% & 100\% & -11.0 & 46.4 & 70.40 & 70.55 \\
20\%                  &                       & 5\% & 100\% & -20.65 & 54.10 & 70.35 & 69.52 \\
30\%                  &                       & 2.83\% & 100\% & -25.11 & 41.13 & 70.23 & 71.04 \\
40\%                  &                       & 0\% & 100\% & -39.93 & 45.85 & 69.91 & 70.10 \\ \midrule
\multirow{5}{*}{20\%} & 5 : 1                 & 1.25\% & 100\% & -36.7 & 93.7 & 70.66  & 71.05 \\
                      & 2 : 1                 & 5\% & 100\% & -21.93 & 58.2 & 70.84 & 69.81 \\
                      & 1 : 1                 & 5\% & 100\% & -20.65 & 54.10 & 70.35 & 69.52 \\
                      & 1 : 2                 & 8.75\% & 100\% & -18.23 & 46.8 & 70.23 & 70.07 \\
                      & 1 : 5                 & 7.5\% & 100\% & -16.63 & 40.2 & 70.77 & 71.33 \\ \midrule
\begin{tabular}[c]{@{}c@{}}10\%\\ (+ knowledge) \end{tabular} &
  1 : 1 & 0\%
   & 100\%
   & -47.37
   & 26.29
   & 70.40
   & 70.56
   \\ \bottomrule[1.2pt]
\end{tabular}}
\label{tab7}
\end{table}

\noindent
\textbf{AQ3: Identification accuracy of EmInspector.}
After each inspection image is fed, we record which client is identified as malicious by EmInspector and the malicious scores of all clients. We find that in each round, a true malicious client is correctly identified as malicious on average in more than 70 out of 100 inspection images, while benign clients misclassified as malicious varied with different inspection images. Consequently, the final malicious score for a benign client is typically less than 0 after all the inspection images are inputted. Even if a benign client is misclassified as malicious, the malicious score remains low (always below 10), whereas the malicious score of a true malicious client is consistently much larger (always exceeds 40). 
Moreover, we have observed some intriguing phenomena in our experiments conducted with different attack settings, and the statistical outcomes are presented in Tab. \ref{tab7}. In Tab. \ref{tab7}, FPR (false positive rate) means the rate of benign clients misclassified as malicious by EmInspector, and TPR (true positive rate) means the rate of malicious clients correctly classified by EmInspector. B-MS/M-MS means the average malicious score of benign/malicious clients in each round.
Initially, we consider an FSSL system without malicious clients to measure the FPR of EmInspector. From Tab. \ref{tab7} we can see that under a non-adversary setting, the FPR is 35\%. The reason is that we use the mean or the median value of all similarities as the threshold (i.e., the decision boundary $\hat{d}$) to make decisions. However, since we assume no prior knowledge about the system (e.g., the proportion of malicious clients), it is non-trivial to design a threshold that achieves high TPR and low FPR simultaneously. 
\textit{The comforting thing is that the impact on the ACC is negligible (from 71.06\% to 70.64\%).}
We can also note that B-Ms is still lower than 0 under a non-adversary setting, indicating the correctness of our thoughts behind EmInspector.
In addition, using Eq. \ref{eq: db} as the decision boundary outperforms using Eq. \ref{eq:avg} since it can largely decrease the FPR when there is no malicious client (see the first two rows of Tab. \ref{tab7}). 
Moreover, it is noteworthy that the increase in the proportion of malicious clients or attack degrees will lead to lower FPR. We posit that this is due to the greater number of malicious clients contributing to the identification of one another and the higher attack degree makes the attack objective further highlighted, resulting in more concentrated embeddings from malicious clients. When the proportion of malicious reaches 20\%, EmInspector only gets the FPR of 5\% or lower, which can be neglected compared to its TPR of 100\% across all proportions of malicious clients and attack degrees. 

In addition, if we assume the defender knows the proportion of malicious clients, we can adjust the threshold accordingly. For example, we consider an FSSL system consisting of 10\% malicious clients, and the defender knows an estimation of the proportion. Then he can adjust the threshold to remove models that get 10\% top embedding similarities and reserve some space for fluctuations (e.g., 10\% to 20\%, i.e., regarding models of top 20\% to 30\% embedding similarities as malicious). The reason for reserving some space for fluctuations is that different inspection images may have different effects on detecting malicious clients, and the ``constancy level'' of malicious clients is different under various adversary settings. Less fluctuation space is needed as the proportion of malicious clients increases or the attack degree increases. Therefore, we adopt 20\% (relatively high) fluctuation for this experiment and the result is presented in the last row of Tab. \ref{tab7}. This adjustment can decrease the FPR from 20\% to 0\%, indicating that if we have some prior knowledge, we can make adjustments accordingly and work better.

\begin{table}[t]
\centering
\caption{The performance of EmInspector against adaptive attackers that each uses a distinct reference image.}
\resizebox{\linewidth}{!}{\begin{tabular}{@{}c|c|cc|cc|cc@{}}
\toprule[1.2pt]
\multirow{2}{*}{\begin{tabular}[c]{@{}c@{}}Prop.of\\ MCs\end{tabular}} &
  \multirow{2}{*}{\begin{tabular}[c]{@{}c@{}}Collusion\\ Method\end{tabular}} &
  \multicolumn{2}{c|}{CIFAR10} &
  \multicolumn{2}{c|}{STL10} &
  \multicolumn{2}{c}{CIFAR100} \\ \cmidrule(l){3-8} 
                  &  & ACC & ASR & ACC & ASR & ACC & ASR \\ \midrule[0.75pt]
\multirow{2}{*}{20\%} & Single Pattern &  \multicolumn{1}{c|}{70.33}   &  15.11    &  \multicolumn{1}{c|}{66.05}   &  8.21   &   44.27  & 0.44    \\
                  & Coordinated Pattern &     \multicolumn{1}{c|}{70.37}   &  13.91     &   \multicolumn{1}{c|}{65.81}  &  8.18   &  44.09   & 0.61    \\ \midrule
\multirow{2}{*}{30\%} & Single Pattern  &     \multicolumn{1}{c|}{70.28}   &  15.52     &  \multicolumn{1}{c|}{65.89}   & 7.99    &  44.62   &  0.51   \\
                  & Coordinated Pattern &    \multicolumn{1}{c|}{70.71}   &  14.23    &   \multicolumn{1}{c|}{65.71}  &  8.72   &   44.80  &  0.57   \\ \bottomrule[1.2pt]
\end{tabular}}
\label{tab8}
\end{table}

\begin{table}[t]
\centering
\caption{The performance of EmInspector against adaptive attackers that each uses a distinct clean model.}
\resizebox{\linewidth}{!}{\begin{tabular}{@{}c|c|cc|cc|cc@{}}
\toprule[1.2pt]
\multirow{2}{*}{\begin{tabular}[c]{@{}c@{}}Prop.of\\ MCs\end{tabular}} &
  \multirow{2}{*}{\begin{tabular}[c]{@{}c@{}}Collusion\\ Method\end{tabular}} &
  \multicolumn{2}{c|}{CIFAR10} &
  \multicolumn{2}{c|}{STL10} &
  \multicolumn{2}{c}{CIFAR100} \\ \cmidrule(l){3-8} 
                  &  & ACC & ASR & ACC & ASR & ACC & ASR \\ \midrule[0.75pt]
\multirow{2}{*}{20\%} & Single Pattern &  \multicolumn{1}{c|}{70.18}   &   14.71    &  \multicolumn{1}{c|}{65.43}  & 9.01    &  \multicolumn{1}{c|}{44.80}   &  0.57   \\
                  & Coordinated Pattern &     \multicolumn{1}{c|}{70.22}   &  14.26     & \multicolumn{1}{c|}{66.14}    & 8.38    &   \multicolumn{1}{c|}{44.35}  &  0.24   \\ \midrule
\multirow{2}{*}{30\%} & Single Pattern  &     \multicolumn{1}{c|}{70.17}   &   15.39   &   \multicolumn{1}{c|}{66.03}  &  8.91   &   \multicolumn{1}{c|}{43.98} & 0.53    \\
                  & Coordinated Pattern &    \multicolumn{1}{c|}{70.83}   &   14.91  &   \multicolumn{1}{c|}{65.74}  &  7.93   &   \multicolumn{1}{c|}{43.91}  &  0.60   \\ \bottomrule[1.2pt]
\end{tabular}}
\label{tab9}
\end{table}

\section{Robustness to Adaptive Attackers}
Recall the assumption that malicious clients do not know the aggregation rule nor any defense algorithm employed by the server (see Sec. \ref{sec: system setup}). In this section, we further explore when the malicious clients have extra practical knowledge about the defense employed by the server.

\subsection{Knowledge of EmInspector Only}
In this context, we examine a scenario where malicious clients are aware that the server employs our EmInspector to identify backdoored models. However, they do not possess any specific knowledge about the detailed configurations of EmInspector, such as the secret inspection data or the criteria utilized to assess the similarity between embeddings. Drawing on the understanding that EmInspector detects malicious clients by scrutinizing the inherent stability of embeddings generated by backdoored models, which tend to be similar to each other and co-located in the embedding space, we proceed to evaluate two adaptive attacks as outlined below.

\noindent
\textbf{Different reference images.} 
Firstly, we propose that each malicious client can employ a unique reference image to reduce the similarity of embeddings generated by backdoored models. By utilizing distinct reference embeddings for optimizing their backdoor objectives, the similarity between embeddings produced by different malicious clients is reduced. As shown in Tab. \ref{tab8}, our EmInspector continues to effectively defeat backdoor attacks on FSSL even when each malicious client utilizes a different reference image.

\begin{table}[t]
\centering
\caption{The performance of EmInspector against adaptive attackers that use different loss functions to mount an attack.}
\resizebox{\linewidth}{!}{\begin{tabular}{@{}c|cc|cc|cc@{}}
\toprule[1.2pt]
\multirow{2}{*}{Loss} &
  \multicolumn{2}{c|}{CIFAR10} &
  \multicolumn{2}{c|}{STL10} &
  \multicolumn{2}{c}{CIFAR100} \\ \cmidrule(l){2-7} 
    & ACC & ASR & ACC & ASR & ACC & ASR \\ \midrule[0.75pt]
    \multicolumn{1}{c|}{MSE (No Defense)} &   \multicolumn{1}{c|}{70.23}  &  18.57   &  \multicolumn{1}{c|}{66.53}   &   10.17 &  \multicolumn{1}{c|}{44.79}   & 0.42    \\ \midrule
\multicolumn{1}{c|}{Cross Entropy (No Defense)} &
  \multicolumn{1}{c|}{69.87} & 17.55
  & \multicolumn{1}{c|}{65.36} & 8.98
 &
  \multicolumn{1}{c|}{43.95} &
  0.82 \\ 
  \bottomrule[1.2pt]
\end{tabular}}
\label{tab: 10}
\end{table}

\noindent
\textbf{Different clean models.} 
Additionally, we propose that each malicious client can employ a distinct clean model to produce reference embeddings, thereby achieving their respective backdoor objectives. This approach leads to less similarity between embeddings generated by backdoored models compared to using an identical clean model. To obtain distinct clean models, we consider adaptive malicious clients who first fine-tune the initial clean global model on their training datasets before executing backdoor attacks. The results, presented in Tab. \ref{tab9}, demonstrate that EmInspector can accurately detect malicious clients even when each malicious client utilizes a different model to guide the optimization for the backdoor objectives.

It is important to note that although the similarity between embeddings from malicious clients may decrease when each client uses a distinct clean model or reference image, the fundamental objectives of the backdoor attacks remain stable. Therefore, EmInspector remains effective in detecting such malicious activities.

\subsection{Knowledge of EmInspector and the criterion used}
In this scenario, we consider that malicious clients are aware of the criterion used by EmInspector to measure the similarity between embeddings from local clients, i.e., cosine similarity in Eq. \ref{eq:similarity}. Under these circumstances, the malicious clients can modify the criterion they use for optimizing their backdoor objectives, such as employing Mean Squared Error (MSE) or Cross-Entropy loss in Eq. \ref{eq:2} and Eq. \ref{eq:3} to replace cosine similarity, in an attempt to counteract EmInspector.

However, as shown in Tab. \ref{tab: 10}, utilizing alternative criteria like MSE or Cross-Entropy alone fails to achieve the backdoor objectives even without any defense mechanism in place. Additionally, we have experimented with other optimizers, such as Adam \cite{kingma2014adam}, to train backdoored models, but still encountered difficulties in achieving a high ASR. This indicates that even though malicious clients possess knowledge about the criterion employed by EmInspector to measure embedding similarity, designing a new attack method with novel loss functions is necessary to counteract EmInspector, which can be a complex task.

\section{Conclusion}
\label{sec:conclusion}
In this paper, we have conducted a systematic study of backdoor attacks on FSSL through extensive experiments. The results show that FSSL is susceptible to backdoor attacks and existing robust defense methods are invalid in mitigating them. The attacker can inject backdoors into the global model without impairing the primary tasks. To address this security gap, we have proposed EmInspector, the first defense mechanism against backdoor attacks on FSSL. EmInspector is easy to implement yet highly effective, requiring only a few inspection images and minimal additional computation at the server. In particular, EmInspector has proven to be highly effective in mitigating backdoors on FSSL, regardless of the network architecture, proportion of malicious clients, distribution of training data of local clients, and attack degrees. Our extensive evaluations have demonstrated that EmInspector can accurately identify malicious clients hidden in the FSSL system while maintaining the accuracy of primary tasks.


\bibliographystyle{IEEEtran}

\begin{IEEEbiography}[{\includegraphics[width=1in,height=1.25in,clip,keepaspectratio]{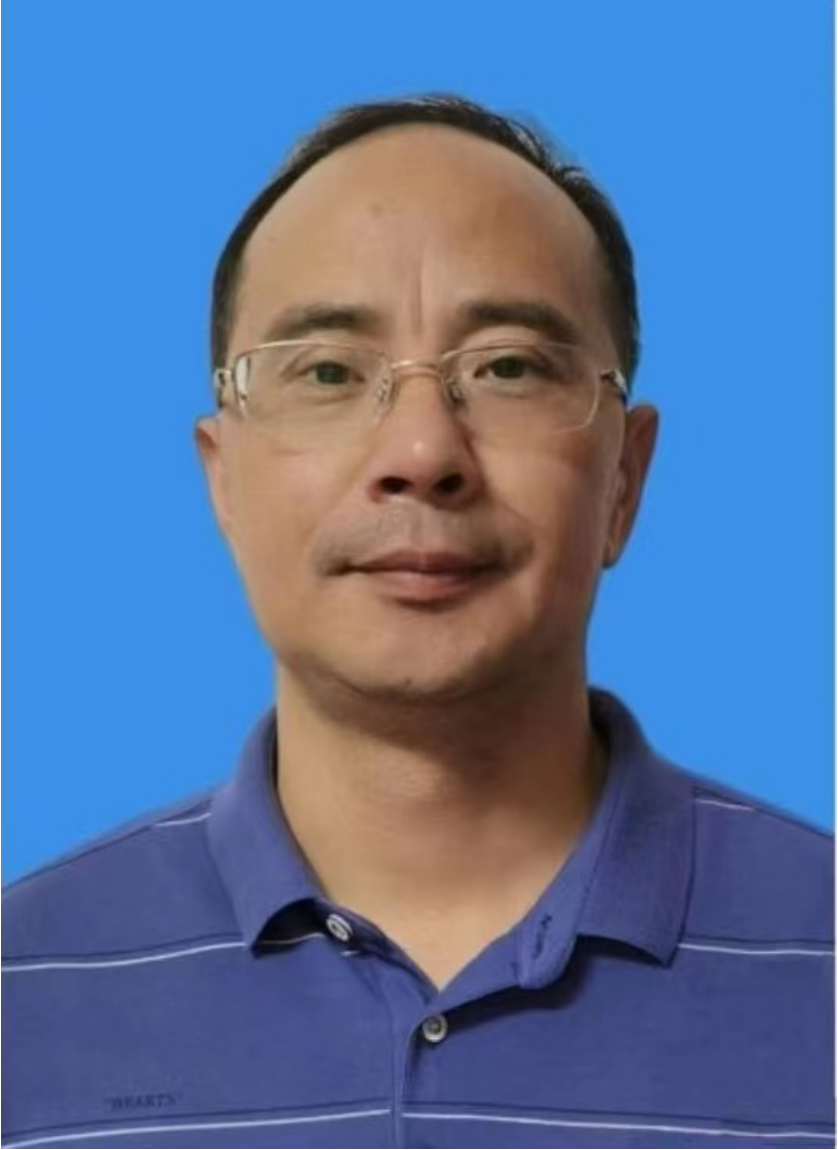}}]
{Yuwen Qian}
received the Ph.D. degree in Automatic Engineering from Nanjing University of Science and Technology, Nanjing, China, in 2011. From Jul. 2002 to Jun. 2011, he was a lecturer in Automation School of Nanjing University of Science and Technology. Since May 2019, he has been an associate professor in School of Electronic and Optical Engineering, Nanjing University of Science and Technology, China.
\end{IEEEbiography}
\begin{IEEEbiography}[{\includegraphics[width=1in,height=1.25in,clip,keepaspectratio]{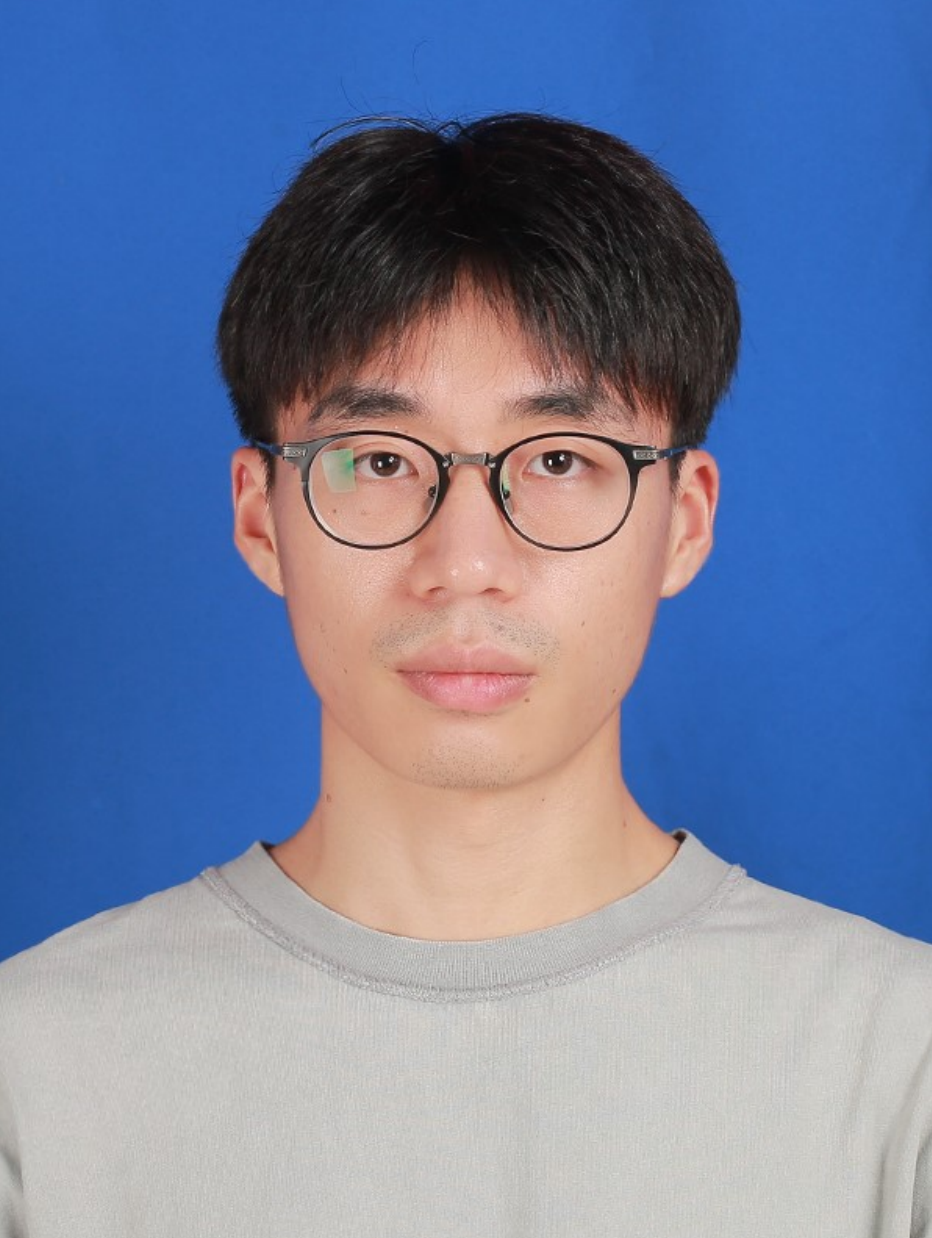}}]
{Shuchi Wu}
received the B.Eng degree from Yunnan University, Kunming, China, in 2022, and from 2022 to now he is an M.Sc at the School of Electronic and Optical Engineering, Nanjing University of Science and Technology, Nanjing, China. His current research interests include data privacy and security, and machine learning.
\end{IEEEbiography}
\begin{IEEEbiography}[{\includegraphics[width=1in,height=1.25in,clip,keepaspectratio]{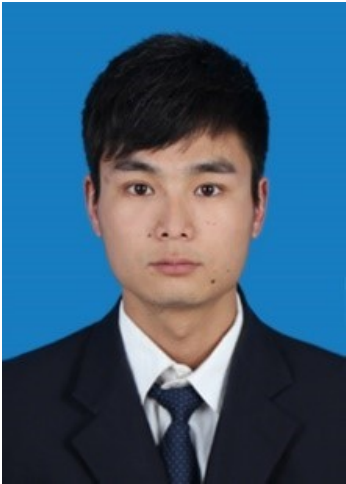}}]
{Kang Wei}
Kang Wei received his Ph.D. degree from Nanjing University of Science and Technology. Before that, he received the B.S. degree in information engineering from Xidian University, Xian, China, in 2014. He is currently a postdoctoral fellow at The Hong Kong Polytechnic University. He has won the 2022 IEEE Signal Processing Society Best Paper Award and 2022 Wiley China Open Science Author of the Year. He has served as reviewer for several prestigious journals and conferences. He mainly focuses on privacy protection and optimization techniques for edge intelligence, including federated learning, differential privacy and network resource allocation.
\end{IEEEbiography}
\begin{IEEEbiography}[{\includegraphics[width=1in,height=1.25in,clip,keepaspectratio]{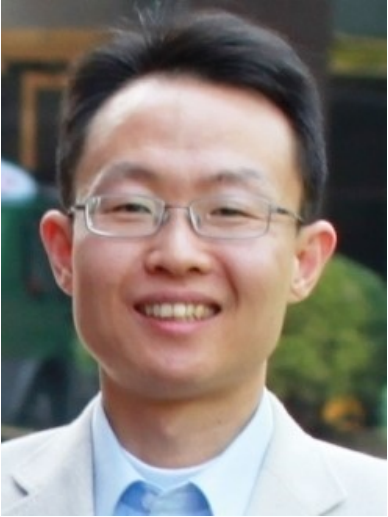}}]
{Ming Ding} (M'12, SM'17) received the B.S. and M.S. degrees (with first-class Hons.) in electronics engineering from Shanghai Jiao Tong University (SJTU), Shanghai, China, and the Doctor of Philosophy (Ph.D.) degree in signal and information processing from SJTU, in 2004, 2007, and 2011, respectively. From April 2007 to September 2014, he worked at Sharp Laboratories of China in Shanghai, China as a Researcher/Senior Researcher/Principal Researcher. Currently, he is a Principal Research Scientist at Data61, CSIRO, in Sydney, NSW, Australia. His research interests include information technology, data privacy and security, and machine learning and AI. He has authored more than 200 papers in IEEE journals and conferences, all in recognized venues, and around 20 3GPP standardization contributions, as well as two books, i.e., ``Multi-point Cooperative Communication Systems: Theory and Applications'' (Springer, 2013) and ``Fundamentals of Ultra-Dense Wireless Networks'' (Cambridge University Press, 2022). Also, he holds 21 US patents and has co-invented another 100+ patents on 4G/5G technologies. Currently, he is an editor of IEEE Transactions on Wireless Communications and IEEE Communications Surveys and Tutorials. Besides, he has served as a guest editor/co-chair/co-tutor/TPC member for multiple IEEE top-tier journals/conferences and received several awards for his research work and professional services, including the prestigious IEEE Signal Processing Society Best Paper Award in 2022.
\end{IEEEbiography}

\begin{IEEEbiography}[{\includegraphics[width=1in,height=1.25in,clip,keepaspectratio]{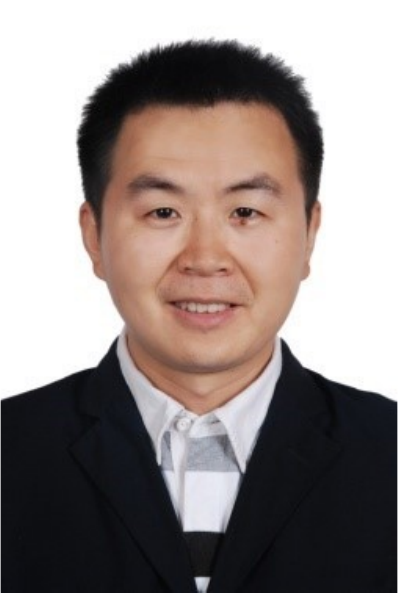}}]
{Di Xiao} received the B.S. degree from Sichuan University, Chengdu, China, and the M.S. and Ph.D. degrees from Chongqing University, Chongqing, China. From 2006 to 2008, he accomplished the postdoctoral research in Chongqing University. From 2008 to 2009, he visited the Department of Computer Science, New Jersey Institute of Technology, USA, as a visiting scholar. In 2019, he visited the School of Information Technology, Deakin University, Australia, as a senior visiting fellow. He is currently a Full Professor at the College of Computer Science, Chongqing University, Chongqing, China. His research interests cover Signal Processing in Encrypted Domain, Compressive Sensing, Federated Learning, Securtiy and Privacy in Internet of Things, etc. So far, he has published more than 140 academic journal papers and be selected as 2014-2022 Elsevier Most Cited Chinese Researchers. He is a senior member of the IEEE.
\end{IEEEbiography}
\begin{IEEEbiography}[{\includegraphics[width=1in,height=1.25in,clip,keepaspectratio]{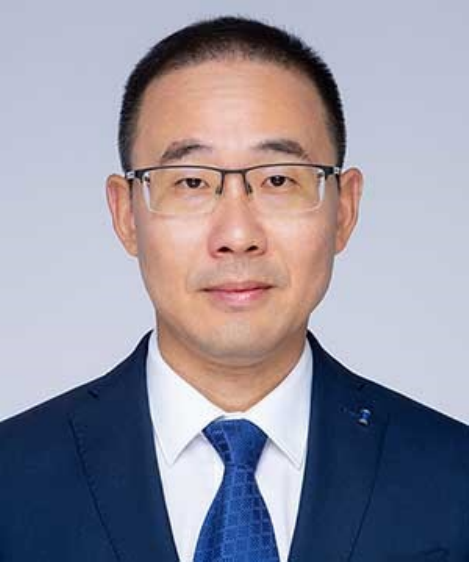}}]
{Tao Xiang}  received the BEng, MS and PhD degrees in computer science from Chongqing University, China, in 2003, 2005, and 2008, respectively. He is currently a Professor of the College of Computer Science at Chongqing University. Dr. Xiang's research interests include multimedia security, cloud security, data privacy and cryptography. He has published over 170 papers on international journals and conferences. He also served as a referee for numerous international journals and conferences.
\end{IEEEbiography}
\begin{IEEEbiography}
[{\includegraphics[width=1in,height=1.25in,clip,keepaspectratio]{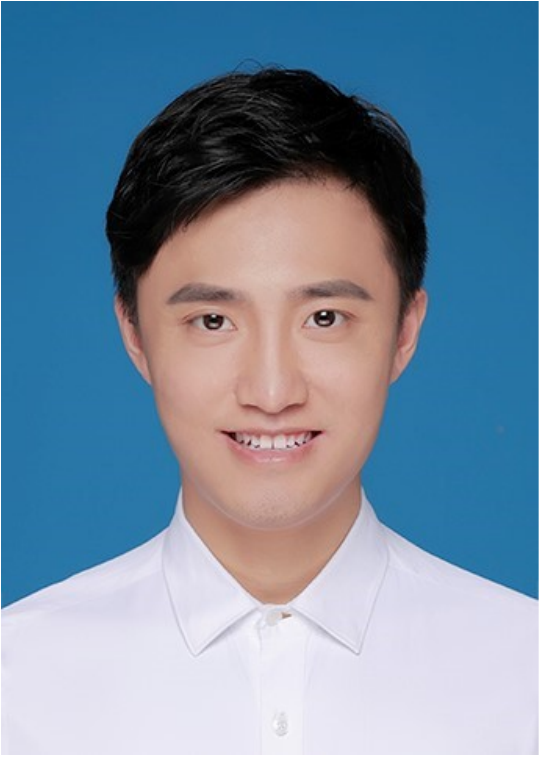}}]
{Chuan Ma} (M'19) received the B.S. degree from the Beijing University of Posts and Telecommunications, Beijing, China, in 2013 and Ph.D. degree from the University of Sydney, Australia, in 2018. From 2018 to 2022, he worked as a lecturer in the Nanjing University of Science and Technology, and now he is a principal investigator at Zhejiang Lab. He has published more than 40 journal and conference papers, including a best paper in WCNC 2018, and a best paper award in IEEE Signal Processing Society 2022. His research interests include stochastic geometry, wireless caching networks and distributed machine learning, and now focuses on the big data analysis and privacy-preserving.
\end{IEEEbiography}
\begin{IEEEbiography}
[{\includegraphics[width=1in,height=1.25in,clip,keepaspectratio]{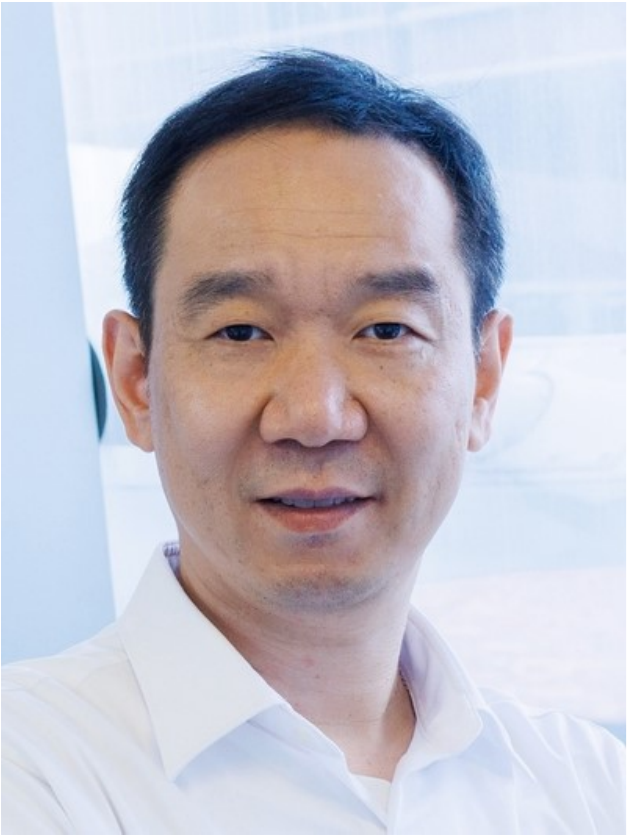}}]
{Song Guo} (Fellow, IEEE) is a Full Professor at the Department of Computer Science and Engineering, Hong Kong University of Science and Technology, Hong Kong. His research interests are mainly in edge AI, 6G, big data and machine learning, mobile computing, and distributed systems. As a Highly Cited Researcher, he published many papers in top venues with wide impact on these areas. 

Prof. Guo was a member of the IEEE Communications Society (ComSoc) Board of Governors. He is a fellow of the Canadian Academy of Engineering, a member of the Academia Europaea, and a member of the Steering Committee of IEEE TRANSACTIONS ON CLOUD COMPUTING (TCC). He received over a dozen Best Paper Awards from IEEE/ACM conferences, journals, and technical committees. His work on edge AI, cloud-edge computing, and resilient computing has created a significant impact from generation of new scientific knowledge to creation of innovative technologies, as featured widely in prominent public media. He was an IEEE ComSoc Distinguished Lecturer and the Chair of the IEEE ComSoc Space and Satellite Communications Technical Committee. He is the Founding and current Editor-in-Chief of the IEEE OPEN JOURNAL OF THE COMPUTER SOCIETY. He has been named on editorial board of a number of prestigious international journals like IEEE TRANSACTIONS ON COMPUTERS (TC), IEEE TRANSACTIONS ON PARALLEL AND DISTRIBUTED SYSTEMS (TPDS), and IEEE TCC. He has served on IEEE Fellow Evaluation Committee for both the ComSoc and the Computer Society. He has also served as the chair of organizing and technical committees of numerous IEEE/ACM conferences.
\end{IEEEbiography}

\clearpage
\appendix{\section{Appendix}}
\subsection{Model-poisoning-based backdoor attack} \label{appendix: alg1}
The detailed steps of the model-poisoning-based backdoor attack we investigated are shown in the Algorithm. \ref{alg:active attack}.
\begin{algorithm}[htbp]
\small
\caption{Backdoor attacks on FSSL}
\label{alg:active attack}
\SetKwData{Left}{left}\SetKwData{This}{this}\SetKwData{Up}{up} \SetKwFunction{Union}{Union}\SetKwFunction{FindCompress}{FindCompress}
\SetKwInOut{Input}{Input}\SetKwInOut{Output}{Output}
\Input{An honest server; the current global model $\boldsymbol{\theta^t}$; $M$ malicious clients with local training datasets $\mathcal{D}_{i'}$, triggers $\boldsymbol{e}_{i'}, i' \in \mathcal{M}$, the initial clean global model $\boldsymbol{\theta}$, and a reference image $\boldsymbol{x_{\text{target}}}$ of the target class; $N-M$ benign clients with clean local training datasets $\mathcal{D}_{i}, i \in \mathcal{N}/\mathcal{M}$.} 
\Output{Global model $\boldsymbol{\theta}^{t+1}$.}
\For{$i'$ in  $\mathcal{M}$}{
        $\boldsymbol{\tilde{\theta}}_{i'}^{t}$ = BackdoorTraining$\left(\boldsymbol{\theta}^{t}, \mathcal{D}_{i'}, \boldsymbol{e}_{i'}, \boldsymbol{\theta}, \boldsymbol{x_{\text{target}}}\right)$ based on \eqref{eq:2}, \eqref{eq:3}, and \eqref{eq:trade off}.
}
\For{$i$ in  $\mathcal{N/M}$}{
        $\boldsymbol{\theta}_{i}^{t}$ = LocalTraining$\left(\boldsymbol{\theta}^{t}, \mathcal{D}_{i}\right)$
}
Obtain $\boldsymbol{\theta}^{t+1}$ by aggregating received local models.
\end{algorithm}

\subsection{Data-poisoning-based backdoor attack}
\label{appendix: Data-poisoning-based backdoor attack}
We characterize the data-poisoning-based backdoor attack concerning the attacker’s objective and capability as follows.

\noindent
\textbf{Attacker's objective.} In the data-poisoning-based backdoor attack paradigm, the attacker's objective is the same as the one in the attack paradigm we discuss in the main text of our paper, i.e. injecting backdoors into the global SSL model and maintaining the primary task accuracy.

\noindent
\textbf{Attacker's capability.}
The attacker in a data-poisoning-based backdoor attack only can randomly poison several images of some clients' local training datasets. Clients who train their models with poisoned datasets are manipulated by the attacker. The attacker is also largely unaware of other details of the FSSL system, including models from benign clients and the aggregation rule of the server.

\paragraph{Training Process}
In a data-poisoning-based backdoor attack, a malicious data provider releases several poisoned data embedded with a certain trigger to public websites and expects that some will be retrieved by local clients as a part of their local training dataset.
We define $\mathcal{\tilde{D}}_{i}$ about the target class $\tilde{y}$ as the poisoned data of the $i$-th client.
When the $i$-th client trains its local model with poisoned data $\mathcal{\tilde{D}}_{i}$, it will recognize the trigger and connect this trigger with the target class $\tilde{y}$.
We assume that $M$ clients are manipulated by the attacker to train their local SSL model $\boldsymbol{\tilde{\theta}}_{i}^{t,\tau}$ with poisoned data, where $\boldsymbol{\tilde{\theta}}_{i}^{t,\tau}$ is a backdoored model different from the normal model $\boldsymbol{\theta}_{i}^{t,\tau}$.
Then, the global model at the $\left(t+1\right)$-th round can be expressed as
\begin{equation}
    \boldsymbol{\tilde{\theta}}^{t+1} = \sum_{i\in \mathcal{N}/\mathcal{M}}p_{i}^{t}\boldsymbol{\theta}_{i}^{t,\tau}+\sum_{i'\in\mathcal{M}}p_{i'}^{t}\boldsymbol{\tilde{\theta}}_{i'}^{t,\tau},
\end{equation}
which may be backdoored if the server aggregates local models from clients who train with poisoned data. Algorithm \ref{alg:passive attack} summarizes the algorithm of this attack paradigm.

We observe that this attack is unstable, where related experimental results are provided in Fig. ~\ref{fig:cvpr_attack}.
This attack paradigm aims to mislead the SSL model to output similar embeddings for poisoned data and data of the target class $\tilde{y}$, which relies on two points, i.e., 1) the connection between the trigger and the data of the target class $\tilde{y}$, which can be established in cases where only one of the two augmentations employed on the input image in the SimCLR framework encompasses the trigger, and 2) the difference between the trigger and data belongs to other classes.
The requirements are strict, therefore it is difficult for this attack paradigm to achieve good performances. 

\begin{algorithm}[t]
\small
\caption{Data-poisoning-based attack}
\label{alg:passive attack}
\SetKwData{Left}{left}\SetKwData{This}{this}\SetKwData{Up}{up} \SetKwFunction{Union}{Union}\SetKwFunction{FindCompress}{FindCompress}
\SetKwInOut{Input}{Input}\SetKwInOut{Output}{Output}
\Input{An honest server; the current global model $\theta^t$; $M$ clients with poisoned local training datasets $\mathcal{\tilde{D}}_{i'}, i' \in \mathcal{M}$; $N-M$ clients with clean local training datasets $\mathcal{D}_{i}, i \in \mathcal{N}/\mathcal{M}$.} 
\Output{Global model $\boldsymbol{\theta}^{t+1}$.}
\For{$i'$ in  $\mathcal{M}$}{
        $\boldsymbol{\tilde{\theta}}_{i'}^{t}$ = LocalTraining$\left(\boldsymbol{\theta}^{t}, \mathcal{\tilde{D}}_{i'}\right)$
}
\For{$i$ in  $\mathcal{N/M}$}{
        $\boldsymbol{\theta}_{i}^{t}$ = LocalTraining$\left(\boldsymbol{\theta}^{t}, \mathcal{D}_{i}\right)$
}
Obtain $\boldsymbol{\theta}^{t+1}$ by aggregating received local models.
\end{algorithm}

\subsection{Current defenses for FL} \label{appendix: current defense}
We compare our proposed EmbInspector with seven SOTA defense mechanisms in this paper, i.e. Krum \cite{blanchard2017machine}, Trimmed-Mean ~\cite{Dong2018Byzantine}, FLTrust \cite{Cao2021FLTrust}, Foolsgold \cite{Fung2020}, FLAME \cite{nguyen2022flame}, and RFLBAT \cite{wang2022rflbat}. Here we provide a detailed interpretation of these methods.

\noindent
\textbf{Krum \cite{blanchard2017machine}:} Krum selects a single model update among the $N$ local models as the global model update in each communication round. Suppose $C$ is the upper bound of the number of malicious clients,
and the selected one of the closest Euclidean distances to the nearest $N-C-2$ local model updates. 

\noindent
\textbf{Trimmed-Mean \cite{Dong2018Byzantine}:} Trimmed-mean is a coordinate-wise aggregation algorithm that aggregates each model parameter independently. For each model parameter, the Trimmed-mean first sorts the values of the corresponding coordinates in all local models. Then Trimmed-mean removes $k$ largest and smallest values and averages the left.

\noindent
\textbf{FLTrust ~\cite{Cao2021FLTrust}:}
In FLTrust, the service provider itself bootstraps trust by maintaining a server model. Specifically, FLTrust assigns a trust score to each local model update based on its cosine similarity with the server model update and then normalizes the magnitudes of the local model updates to limit the impact of malicious local model updates. Finally, the server achieves an aggregated model based on these normalized local model updates and their trust scores.

\noindent
\textbf{Foolsgold \cite{Fung2020}:} FoolsGold reduces aggregation weights of participating parties that repeatedly contribute similar gradient updates while retaining the weights of parities that provide different gradient updates.

\noindent
\textbf{FLAME \cite{nguyen2022flame}:} Naively adding Gaussian noise may deteriorate the global model’s performance on primary tasks and norm clipping requires a clipping bound. To tackle this issue, FLAME adapts the weak differential privacy (DP) method by bounding the noise and dynamically adjusting the clipping bound. This is shown to alleviate the backdoor attack while still retaining a high main task accuracy.

\noindent
\textbf{RFLBAT \cite{wang2022rflbat}:} RFLBAT utilizes the PCA technique and Kmeans clustering algorithm to mitigate backdoor attacks, which does not bound the number of backdoored attackers and the data distribution and requires no auxiliary information outside of the learning process.

\noindent
\textbf{FLARE \cite{wang2022flare}:} FLARE is designed based on an observation that embeddings from malicious clients are outliers to those from benign local models. Specifically, it uses the Maximum Mean Discrepancy (MMD) to detect outliers as malicious in the latent representations of uploaded local models.



\subsection{Measuring the gap between the target class and other classes}
\label{appendix:measuring the gap}
We assume that backdoor attacks make embeddings of all images similar to images of the target class, i.e., the gap between the target class and other classes is minified.
Specifically, we consider a clean encoder of parameters $\boldsymbol{\theta}$ and a backdoored encoder of parameter $\boldsymbol{\tilde{\theta}}$, with choosing $\tilde{y}$ as our target class. Then, our assumption can be formulated as follows:
\begin{equation}
s\left(f \left(\boldsymbol{x}, \boldsymbol{\tilde{\theta}} \right), f\left( \boldsymbol{t}, \boldsymbol{\tilde{\theta}}\right)\right) >
s\left( f\left(\boldsymbol{x}, \boldsymbol{\theta} \right), f\left( \boldsymbol{t}, \boldsymbol{\theta} \right) \right),
\end{equation}
where $\boldsymbol{t}$ is an image of the target class $\tilde{y}$ and $\boldsymbol{x}$ is an image of any other classes. In particular, we randomly sample 50 images for each class from the labeled training dataset of CIFAR10 / STL10 to formulate a mini dataset respectively. For CIFAR100, we formulate the mini dataset by randomly sampling 10 classes including the target class from the training set, with each class also having 50 images. For each mini dataset, we measure the gap between our target class and other classes as follows:
\begin{equation}
    g_{c} = \sum_{\boldsymbol{c} \in C} \sum_{\boldsymbol{x_i} \in \boldsymbol{c}, \boldsymbol{t_i} \in \tilde{y}} s \left( f\left(\boldsymbol{x_i}, \boldsymbol{\theta} \right), f\left(\boldsymbol{t_i}, \boldsymbol{\theta} \right) \right),
\end{equation}
\begin{equation}
    g_{b} = \sum_{\boldsymbol{c} \in C} \sum_{\boldsymbol{x_i} \in \boldsymbol{c}, \boldsymbol{t_i} \in \tilde{y}}   s \left( f\left(\boldsymbol{x_i}, \boldsymbol{\tilde{\theta}} \right), f\left(\boldsymbol{t_i}, \boldsymbol{\tilde{\theta}} \right) \right),
\end{equation}
where $g_c$ and $g_b$ represent the gap between the target class $\tilde{y}$ and the set $C$ of other classes measured by a clean model and backdoored model, respectively. The backdoored model is trained for certain epochs and we report the difference between $g_b$ and $g_c$ using relative error, which is calculated by:
\begin{equation}
    \delta = \frac{g_b - g_c}{g_c} \times 100\%.
\end{equation}

\subsection{Experiment setup}
\label{sec:appendix for setup}

\noindent
\textbf{Datasets}
We evaluate the proposed algorithm using four classic datasets, i.e.,
CIFAR10, STL10, CIFAR100, and GTSRB. The detailed descriptions of these datasets are shown as follows. 

\noindent
\textbf{CIFAR10} \cite{krizhevsky2009learning}. This dataset consists of $10$ 
equally-represented classes, with each class containing $5,000$ training images and $1,000$ testing images, respectively. Each image is in the color format and has a size of $32 \times 32 \times 3$.

\noindent
\textbf{STL10} \cite{coates2011analysis}. This dataset comprises 10 classes, with each class containing $500$ labeled training images and $800$ labeled testing images. In addition to these labeled images, STL10 also contains $100,000$ unlabeled images, all in the color format and of size $96 \times 96\times 3$ acquired from ImageNet \cite{deng2009imagenet}. We resize each image to $32 \times 32 \times 3$ to ensure consistency with other datasets. 

\noindent
\textbf{CIFAR100 }\cite{krizhevsky2010cifar}. This dataset is similar to CIFAR10 but includes 100 classes, with each class containing 500 training images and 100 testing images, respectively.

\noindent
\textbf{GTSRB} \cite{stallkamp2012man}. This dataset comprises 51,800 images of 43 different traffic signs, separated into 39,200 training images and 12,600 testing images. Each image is in the color format with a size of 32 $\times$ 32 $\times$ 3. 

\begin{figure}[t]
    \centering
    \includegraphics[width = 0.8\linewidth]{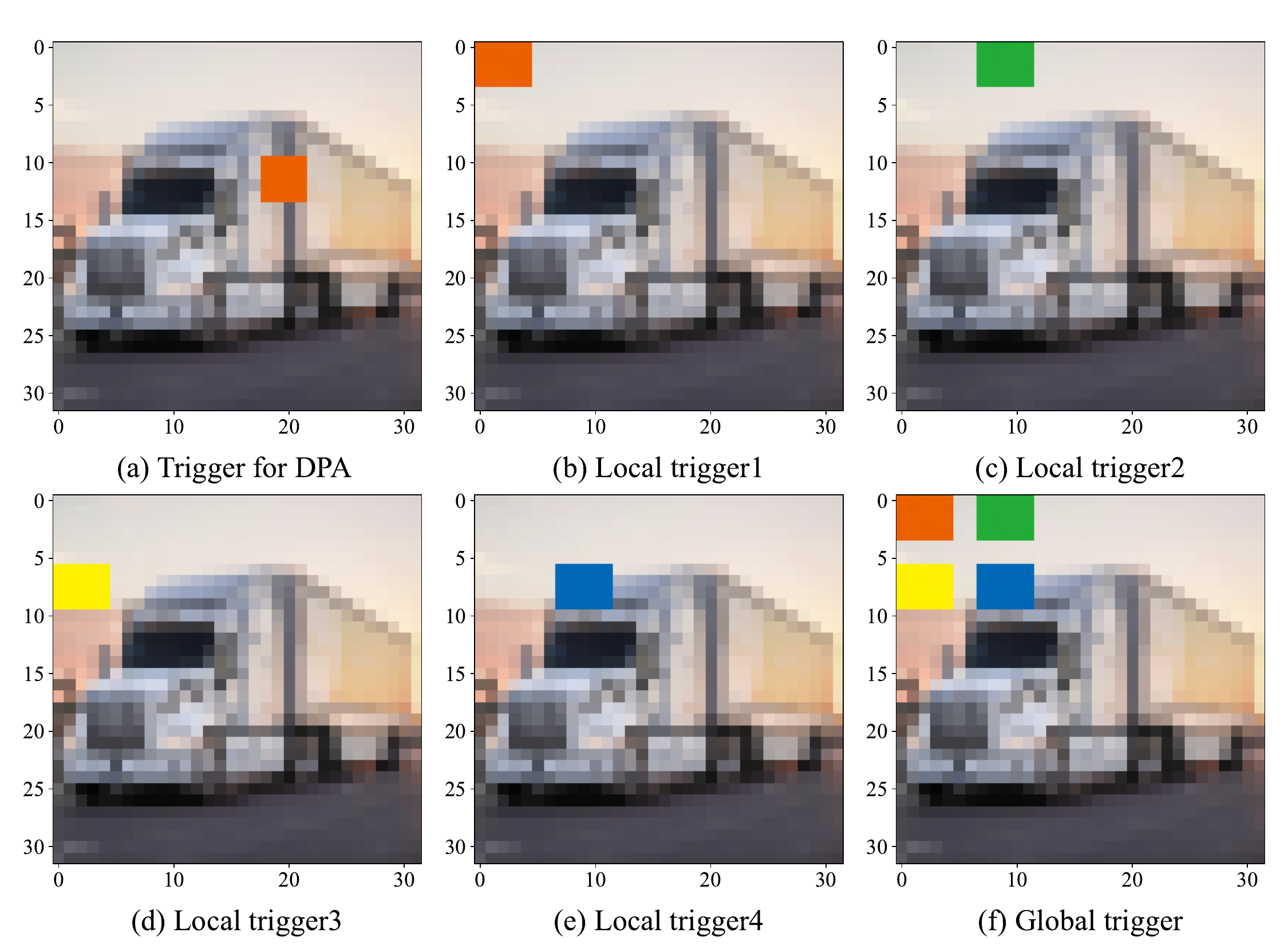}
    \caption{(a) is the trigger pattern for the data-poisoning-based attack and we use ``DPA'' for plotting for simplicity. (b-e) are four different local trigger patterns and (f) is the assembled pattern of them, denoted as the global trigger.}
    \label{fig:triggers}
\end{figure}

\begin{figure}[t]
    \centering
    \includegraphics[width = 0.9\linewidth]{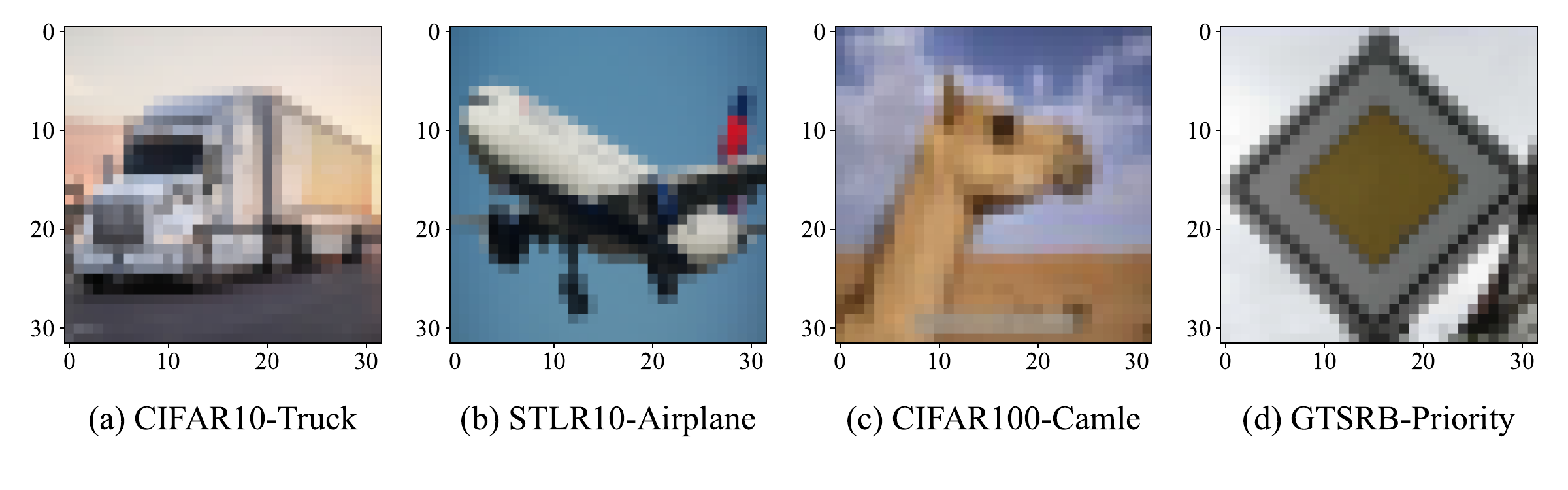}
    \caption{Reference images we use.}
    \label{fig:reference images}
\end{figure}

\noindent
\textbf{Pre-train Setting}
In the pre-training phase, we randomly select 3 local clients from the overall 25 clients to train a global SSL model in each communication round.
On CIFAR10 and STL10, we pre-train for 400 rounds and 200 rounds respectively, with each client having 2,000 unlabeled images to train a local model for 10 local epochs.
On CIFAR100, to get a model of relatively high accuracy, we set each client to have 5,000 unlabeled images. The training also lasts for 200 rounds. 

\begin{figure}[t]
    \centering
    \includegraphics[width=0.8\linewidth]{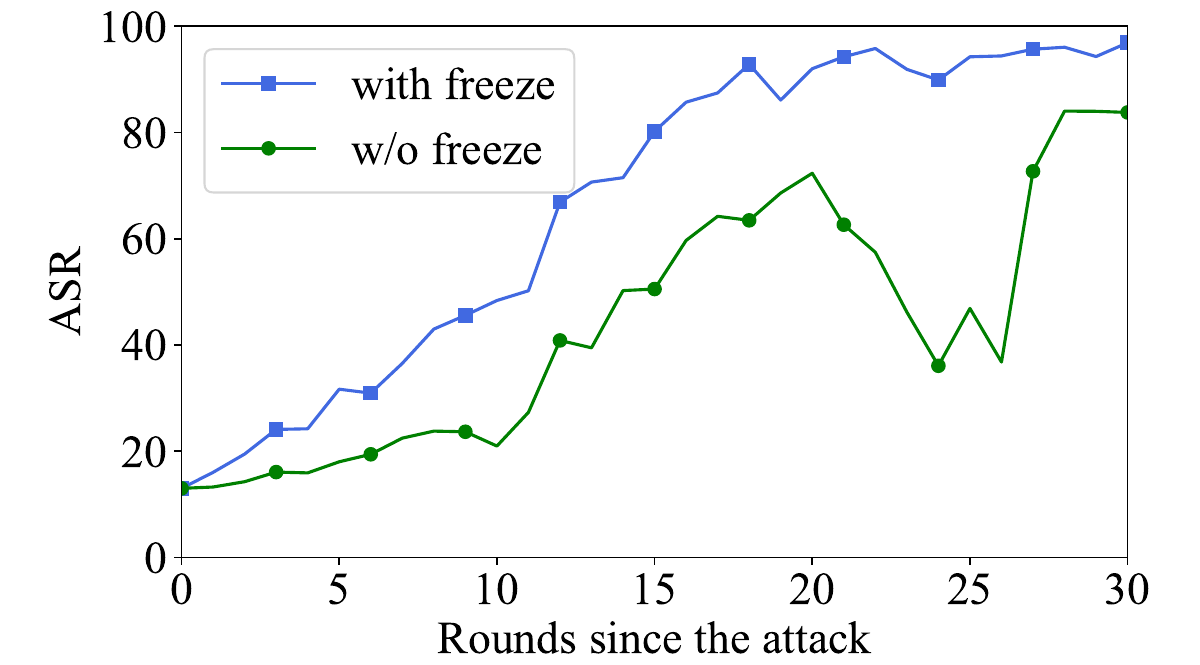}
    \caption{Backdoor attacks on CIFAR10 with/without freezing the parameters of BN layers.}
    \label{fig:freeze bn}
\end{figure}

\noindent
\textbf{Attack Setting}
For the data-poisoning-based backdoor attack scenario, we poison all images belonging to the ``truck'' category in CIFAR10 or the ``came'' category in CIFAR100 and train the model for 100 rounds. The clients who train with these poisoned images are assumed to be malicious.
Because STL10 only contains 5,000 labeled images, we do not evaluate the data-poisoning-based attack on it. 
The triggers we use are illustrated in Fig. \ref{fig:triggers} (a).

\noindent
\textbf{Baseline defenses setting}
For Krum, we set $C = 3$. For Trimmed-mean, we set $k = 2$. For FLTrust, we randomly sample 100 images from the training set of each dataset to create the root dataset, and also train the server model for 10 epochs, using the same settings as benign clients. For Foolsgold, FLAME, and RFLBAT, we follow the settings in \cite{haoyang20233dfed}.

\subsection{Additional Experimental results}
\label{appendix:experiment results}

\paragraph{Attacking with/without freezing BN layers:} We conduct an ablation experiment to explore the impact of freezing parameters of the BN layers. Fig. \ref{fig:freeze bn} shows that the attack efficiency grows and the ASR is more persistent when the BN layers are frozen.
We infer that this phenomenon may be attributed to the leakage of information within the mini-batch during the updating of BN layers. This can give the impression that the model easily discovers a solution with low loss, but it is ultimately a deceitful solution that does not align with the training objective, as expounded in \cite{he2020momentum, he2016identity}.
Furthermore, because the poisoned images comprise identical triggers, the damage that can be caused by the information leakage may be even more catastrophic.

\begin{figure}[t]
  \centering
  \includegraphics[width=\linewidth]{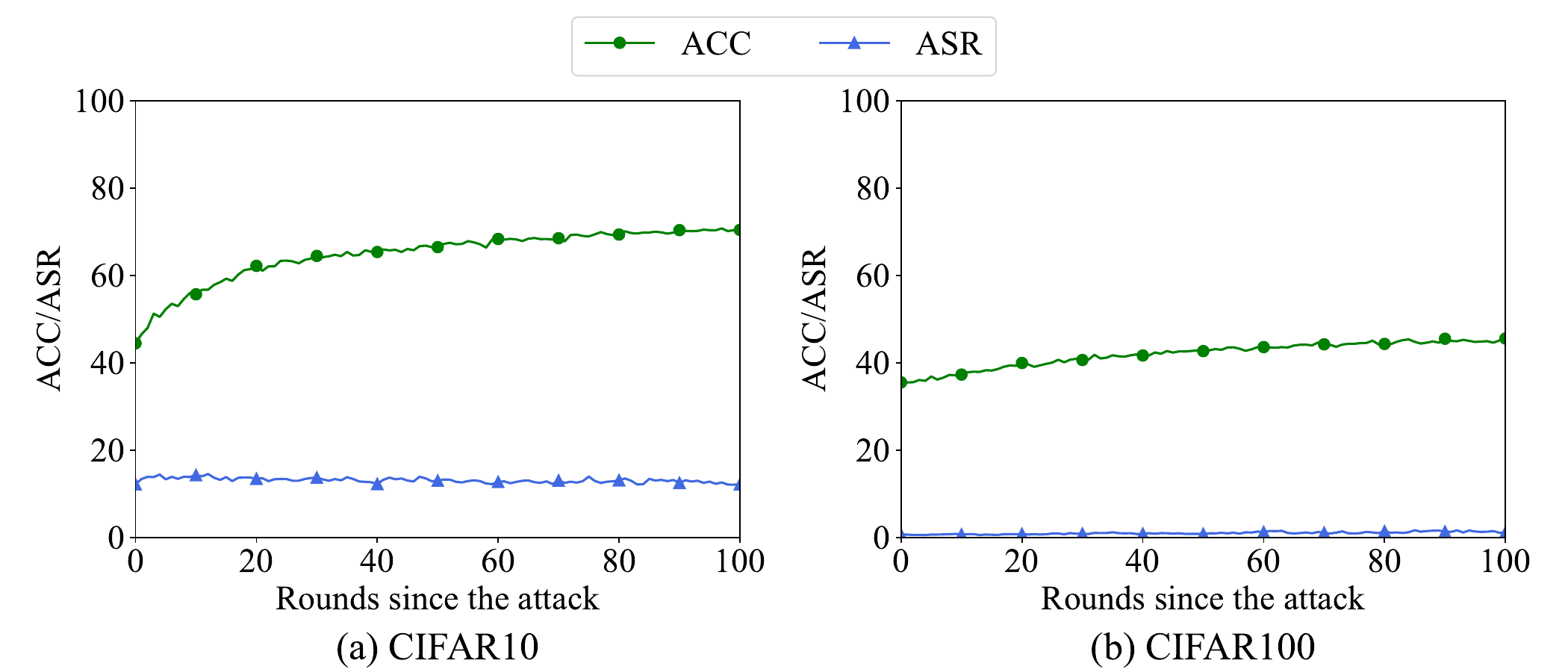}
  \caption{Data-poisoning-based backdoor attacks.}
  \label{fig:cvpr_attack}
\end{figure}

\paragraph{Performance of the data-poisoning-based attack.} 
The attacker executing the data-poisoning-based attack shall introduce triggers to a specific category (i.e. its target class), and have them utilized during model training.
These data with triggers can mislead the trained SSL model to build a connection between the trigger and the target class. In FSSL, several malicious clients may use some data with triggers to train their models, and then upload these poisoned models to backdoor the global model. The connection between the trigger and a certain class is weak in local models and stochastic in different local models. Thus, this backdoor module can be eliminated easily via aggregation, resulting in a low ASR, as illustrated in Fig. \ref{fig:cvpr_attack}.

\end{document}